\documentclass[preprint2,times,tighten]{aastex62}
\pdfoutput=1 
\usepackage{amsmath,amstext,amssymb}
\usepackage{wasysym} 						
\usepackage{verbatim}
\usepackage{soul}
\usepackage[T1]{fontenc}
\usepackage{apjfonts} 
\usepackage[figure,figure*]{hypcap}
\usepackage{rotating}
\usepackage{mathtools}
\usepackage{multirow}

\shorttitle{Modeling Turbulent Material in the CGM}
\shortauthors{Buie et al.}

\begin{document}

\title{Modeling Photoionized Turbulent Material in {the Circumgalactic Medium II: Effect of Turbulence within a Stratified Medium}}

\author{Edward Buie II}
\affiliation{Arizona State University School of Earth and Space Exploration, P.O. Box 871404, Tempe, AZ 
85287, USA}
\author{William J. Gray}
\affiliation{CLASP, College of Engineering, University of Michigan, 2455 Hayward St., Ann Arbor, MI 48109, USA}
\author{Evan Scannapieco} 
\affiliation{Arizona State University School of Earth and Space Exploration, P.O. Box 871404, Tempe, AZ 
85287, USA}
\author{Mohammadtaher Safarzadeh} 
\affiliation{Department of Astronomy and Astrophysics, University of California, Santa Cruz, CA 95064, USA}

\begin{abstract}
The circumgalactic medium (CGM) of nearby star-forming galaxies shows clear indications of \ion{O}{6} absorption accompanied by little to no detectable \ion{N}{5} absorption. This unusual spectral signature, accompanied by highly non-uniform absorption from lower ionization state species, indicates that the CGM must be viewed as a dynamic, multiphase medium, such as occurs in the presence of turbulence. Motivated by previous isotropic turbulent simulations, we carry out chemodynamical simulations of stratified media in a Navarro-Frenk-White (NFW) gravitational potential with a total mass of $10^{12}$~M$_{\odot}$ {and turbulence that decreases radially.} The simulations assume a metallicity of 0.3~Z$_{\odot}$, a redshift zero metagalatic UV background, and they track ionizations, recombinations, and species-by-species radiative cooling using the MAIHEM package. We compare a suite of ionic column densities with the COS-Halos sample of low-redshift star-forming galaxies. Turbulence with an average one-dimensional velocity dispersion $\approx 40$~km~s$^{-1}$, corresponding to an energy injection rate of $\approx 4 \times 10^{49}$~erg~yr$^{-1},$ produces a CGM that matches many of the observed ionic column densities and ratios.  In this simulation, the $N_{\rm N\ V}/N_{\rm O\ VI}$ ratio is suppressed from its equilibrium value due to a combination of radiative cooling and cooling from turbulent mixing. This level of turbulence is consistent with expectations from observations of better constrained, higher-mass systems, and could be sustained by energy input from supernovae, gas inflows, and dynamical friction from dark matter subhalos. We also conduct a higher resolution $\approx 40$~km~s$^{-1}$ run which yields smaller-scale structures, but remains in agreement with observations.
\end{abstract}


\section{Introduction} \label{intro}

The circumgalactic medium (CGM) {is a cloud of diffuse baryons that resides in the gravitational potential of a host galaxy. This medium extends hundreds of kpc, and it plays an essential role in facilitating the host galaxy's evolution through the transfer of baryons in} processes such as accretion, galactic winds, and feedback from active galactic nuclei (AGN) \citep[e.g.][]{2013lilly,2015voit,2015crighton,fox2017gas,2017muratov,2017reviewARA&A..55..389T}.  {Unfortunately, its diffuse nature makes this medium difficult to observe through emission.}

Instead, our knowledge of the CGM relies on observations of absorption in the spectra of background quasars. Much of the earliest measurements of this type made use of the WFPC-2 camera on the \textit{Hubble Space Telescope (HST)} as well as the HIRES spectrograph on the Keck telescopes, which allowed the capability for observations of distant, $z \approx 3$ systems. \citet{1998steidel} provides a summary of these early findings, which attempted to understand the relationship between galaxies,  the CGM, and the intergalactic medium (IGM).

To gain further insight into this relation, the Keck Baryonic Structure Survey \citep[KBSS;][]{rudie2012gaseous} {was undertaken to observe galaxies during the cosmic peak of star-formation ($z \approx 2 - 3$).} {Theory tells us that these higher-redshift systems should be actively accreting material through cold filamentary structures to fuel the high star formation rates observed \citep[e.g.][]{birnboim2003virial,ocvirk2008bimodal,brooks2009role,faucher2011small}.} {\citet{rudie2012gaseous} revealed this cooler material in \ion{H}{1} absorption as well as an anti-correlation between \ion{H}{1} absorbers and transverse distance to the host galaxy, otherwise known as the impact parameter. 

At low resdshift, the CGM has been probed with multiple generations of HST instruments.  Using the Faint Object Spectrograph (FOS), \cite{Chen1998,Chen2001a,Chen2001b} was able to derive constraints on high equivalent width \ion{H}{1} and \ion{C}{4} absorbers in the CGM around $z<1$ galaxies. More recently, the  Cosmic Origins Spectrograph (COS) has given us increased sensitivity to the diffuse absorption  \citep{2009AIPC.1135..301S}. Using this instrument, \citet{tuml2013ApJ...777...59T} observed the CGM of $z$ $\lesssim$ 0.5, $M_{*} = 10^{9.5} - 10^{11.5}$~M$_{\odot}$ galaxies out to an impact parameter of $b = 150$~kpc in the COS-Halos survey. The survey showed a neutral H component associated with nearly every galaxy, along with large amounts of \ion{O}{6} absorption in the CGM of star-forming galaxies that was not observed in the non star-forming sample.

\citet{2013werkApJS..204...17W} {and \citet[][hereafter W16]{werk2016ApJ...833...54W} expanded on the COS-Halos survey and found absorption from singly ionized species resulting from a cooler} ($T \approx 10^4-10^5$~K) CGM phase. This {absorption was found to decrease with impact parameter similarly to earlier trends observed in \ion{H}{1}.} {Furthermore, W16 found \ion{O}{6} absorption in sightlines at a variety of impact parameters and used them to trace the hot ($T \gtrsim 10^{5.5}$) ambient medium. \ion{N}{5} absorption, however, was largely absent, an odd finding as the ionization potentials for these ions differs by only $\approx 30\%.$

W16 also {included examinations of a variety of models to explain these findings \citep{2007gnatApJS..168..213G,2009gnatApJ...693.1514G,2012wakkerApJ...749..157W,2017cloudyRMxAA..53..385F}, but, to fit the data,these either required unphysical conditions for the gas or a narrow range in parameter space.  Motivating by these issues, in \citet[][hereafter Paper I]{buie2018modeling} we undertook a comprehensive Fai investigation of the impact of homogeneous turbulence on the properties of the CGM.}

Several CGM processes that are likely to drive significant turbulence. Theoretical work supported by observations shows distinctly cooler material flowing into star-forming galaxies \citep{ 2005coolflowMNRAS.363....2K, 2006coolflowMNRAS.368....2D, 2009coolflowApJ...700L...1K, 2011coolflowApJ...738...39S,2012inflowApJ...747L..26R}. This cooler material will drive turbulence along the boundaries between it and the hotter ambient halo, and possibly throughout the entire medium.

Many galaxies also display global outflows, which inject material, momentum, and energy into the surrounding medium. These outflows are typically driven by  supernovae resulting from ongoing star-formation  \citep[][for a review]{veilleux2005galactic}. Observations of this outflowing gas reveal complex multiphase features with $\gamma$ and X-ray emission from 10$^{8}$--10$^{7}$~K plasma \citep{strickland2007iron,strickland2009supernova,laha2018study} and UV absorption from 10$^{6}$--10$^{4}$~K material \citep{rubin2014evidence,fox2015probing,heckman2017cos}. 

Finally, dynamical friction resulting from the gravitational interactions between the CGM and dark matter subhalos may also lead to turbulent motions, as gas is accelerated in the wake of moving gravitational potentials \cite{Chandrasekhar1943}. As this material loses angular momentum and energy, it will merge with its host halo, although the merging timescales for these events depend on treatments of the orbital energy, angular momentum, and subhalo mass \citep[e.g.][]{kauffmann1993formation,somerville1999semi,boylan2008dynamical}.

In Paper I, we modeled the non-equilibrium chemical evolution of homogeneous turbulent media exposed to the extragalactic UV background (EUVB). We found that a one-dimensional (1D) velocity dispersion of $\sigma_{\rm 1D} \approx 60$~km~s$^{-1}$ replicated many of the observed features within the CGM, such as clumping of low ionization-state ions and the existence of \ion{O}{6} at moderate ionization parameters. However, unlike observations, \ion{N}{5} arose in our simulations with derived column densities of a similar magnitude to those of \ion{O}{6}. We also found that increasing the turbulent velocity dispersion led to a thermal runaway, resulting in a media not representative of the CGM around star-forming galaxies.

To gain a more realistic picture of turbulent systems, here we conduct direct numerical simulations of a stratified turbulent astrophysical media contained within a gravitational potential.  We compare these results directly with observations to determine the extent to which multiphase observations of the CGM can be explained by the presence of sustained, stratified turbulence. 

The paper is organized as follows: in \S \ref{methods} we outline the code used to model the CGM. In \S  \ref{results} we present our results with a focus on \ion{O}{6} and \ion{N}{5} abundances as well as compare our results to W16 and in {\S  \ref{constraints} we discuss the observational constraints on turbulence and the energy requirements for our turbulence while providing comparisons to real drivers of such turbulence.} Finally, in \S  \ref{summary} we give our concluding remarks.

\section{Methods} \label{methods}

\subsection{The MAIHEM Code} \label{maihem}
We used Models of Agitated and Illuminated Hindering and Emitting Media (MAIHEM\footnote{http://maihem.asu.edu/})  to simulate a stratified turbulent CGM. MAIHEM is a three-dimensional (3D) cooling and chemistry package built using FLASH (Version 4.5), an open-source hydrodynamics code \citep{fryxell2000flash}. This package evolves the non-equilibrium chemistry network of 65 ions, including hydrogen (\ion{H}{1} and \ion{H}{2}), helium (\ion{He}{1}--\ion{He}{3}), carbon (\ion{C}{1}--\ion{C}{6}), nitrogen (\ion{N}{1}--\ion{N}{7}), oxygen (\ion{O}{1}--\ion{O}{8}), neon (\ion{Ne}{1}--\ion{Ne}{10}), sodium (\ion{Na}{1}--\ion{Na}{3}), magnesium (\ion{Mg}{1}--\ion{Mg}{4}), silicon (\ion{Si}{1}--\ion{Si}{6}), sulfur (\ion{S}{1}--\ion{S}{5}), calcium (\ion{Ca}{1}--\ion{Ca}{5}), iron (\ion{Fe}{1}--\ion{Fe}{5}), and electrons. This includes solving for dielectric and radiative recombinations, collisional ionizations with electrons, charge transfer reactions, and photoionizations by a UV background. 

This package was first developed in \citet{gray2015atomic} and later improved upon with the inclusion of an ionizing background in \citet{gray2016atomic}. Most recently, in \citet{gray2017effect}, several charge transfer reactions, radiative recombination rates, and dielectronic recombination rates from \citet{aldrovandi1973radiative,shull1982ionization,arnaud1985updated} and 
photoionizing and photoheating rates from \citet{verner1995analytic} and \citet{verner1996atomic} were added and updated in the code. 

The equations solved by MAIHEM are given in \citet{gray2016atomic} and are invariant under the transformation $x \rightarrow \lambda x,\ t \rightarrow \lambda t,\ \rho \rightarrow \rho/\lambda$ meaning the final steady-state abundances depend only on the mean density multiplied by the driving scale of turbulence, $nL$, the one-dimensional (1D) velocity dispersion of the gas, $\sigma_{\rm 1D}$, and the ionizing EUVB radiation which may be parameterized by $U$; the ratio of number of ionizing photons to the number density of hydrogen $n_{\rm H}$, or alternatively,
\begin{equation}
\label{eq:1}
U  \equiv \frac{\Phi}{n_{\rm H} c}, 
\end{equation}
where $\Phi$ is the total photon flux of ionizing photons, and $c$ is the speed of light.

In addition, we used an unsplit solver based on \citet{2013leeJCoPh.243..269L} to solve the hydrodynamic equations.  Here we made use of a hybrid Riemann solver that uses the Harten Lax and van Leer (HLL) solver \citep{einfeldt1991godunov} in places with strong shocks or rarefactions and the Harten--Lax--van Leer--Contact (HLLC) solver \citep{toro1994restoration,tororiemann} in smoother flows to stabilize the code as turbulence ensues. We refer the reader to \citet{gray2015atomic} and \citet{gray2016atomic} for further details. 

\subsection{Turbulence and Gravity} \label{turb_gravity}

We modeled turbulence driven through solenoidal modes ($\nabla \cdot F = 0$) \citep{pan2010}. 
These modes depend on the driving scale of turbulence and undergo a direct Fourier transformation to the physical space and are added to the acceleration. The acceleration was varied with radii as
\begin{equation}
a_{x,y,z} = a^0_{x,y,z} \left(\frac{r + 0.3 R_{\rm vir}}{0.5 R_{\rm vir}}\right)^{\alpha},
\label{equ:trend}
\end{equation} 
where $a^0_{x,y,z}$ is the original acceleration term resulting from the direct Fourier transform, $r$ is the radius to a cell, $R_{\rm vir}$ is the virial radius of the halo, and $\alpha$ is a dimensionless parameter that controls how the acceleration behaves with radius. We set $\alpha=-1$ such that stirring was strongest towards the center and falls off with radius to capture the behavior of star-forming galaxies.

Furthermore, we added a dark matter halo to produce a stratified baryonic matter density profile in these simulations that assumes a
\citep[][hereafter NFW]{Navarro1996} profile under the $\Lambda$CDM model.
\begin{equation}
    \rho(r) = \frac{\rho_0}{\frac{r}{R_s} 
    \left(1 + \frac{r}{R_s} \right)^2},
\end{equation}
where $\rho_0  = M_{\rm halo} \left[4\pi R_s^{3} \left[ \ln(1+c)- \right. \right.$ $\left. \left. c/(1+c) \right] \right]^{-1}$ is the central dark matter density, $M_{\rm halo}$ is the mass of the halo, $R_s = R_{\rm vir}/c$ is the scale radius, and $c$ is the concentration parameter of the halo.

\begin{figure}[t]
    \centering
    \includegraphics[width=1.0\linewidth]{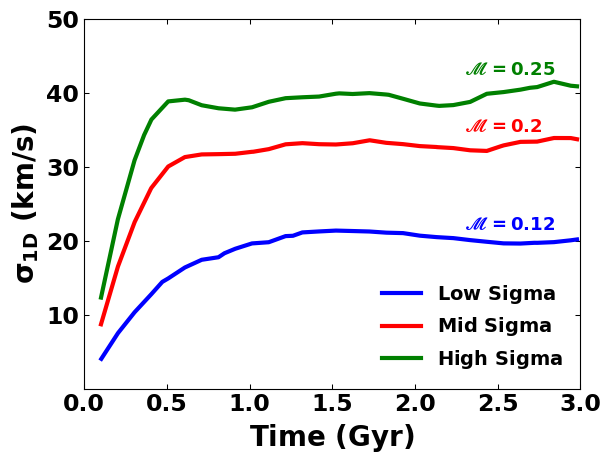}
    \caption{1D velocity dispersion ($\sigma_{\rm 1D}$) vs. time for the Low (blue), Mid (red), and High (green) runs. We also show the 1D Mach number, $\mathcal{M}$, at 3~Gyrs for the runs in the same colors as their $\sigma_{\rm 1D}$ curves.}
    \label{fig:velocity}
\end{figure}

This dark matter halo was also used to generate the gravitational acceleration of the baryonic matter. The cosmological parameters used in this work are from the Planck 2018 Collaboration \citep{planck2018} and are $h$ = 0.674, $\Omega_m$ = 0.315, $\Omega_b$ = 0.049, and $\Omega_{\Lambda}$ = 0.685,  where $h$ is the Hubble constant in units of 100 km s$^{-1}$ Mpc$^{-1}$, and $\Omega_m$, $\Omega_b$, and $\Omega_\Lambda$, are the total matter, baryonic, and vacuum densities, respectively, in units of the critical density.\\

\begin{figure*}[t]
\centering
\includegraphics[width=1.0\linewidth]{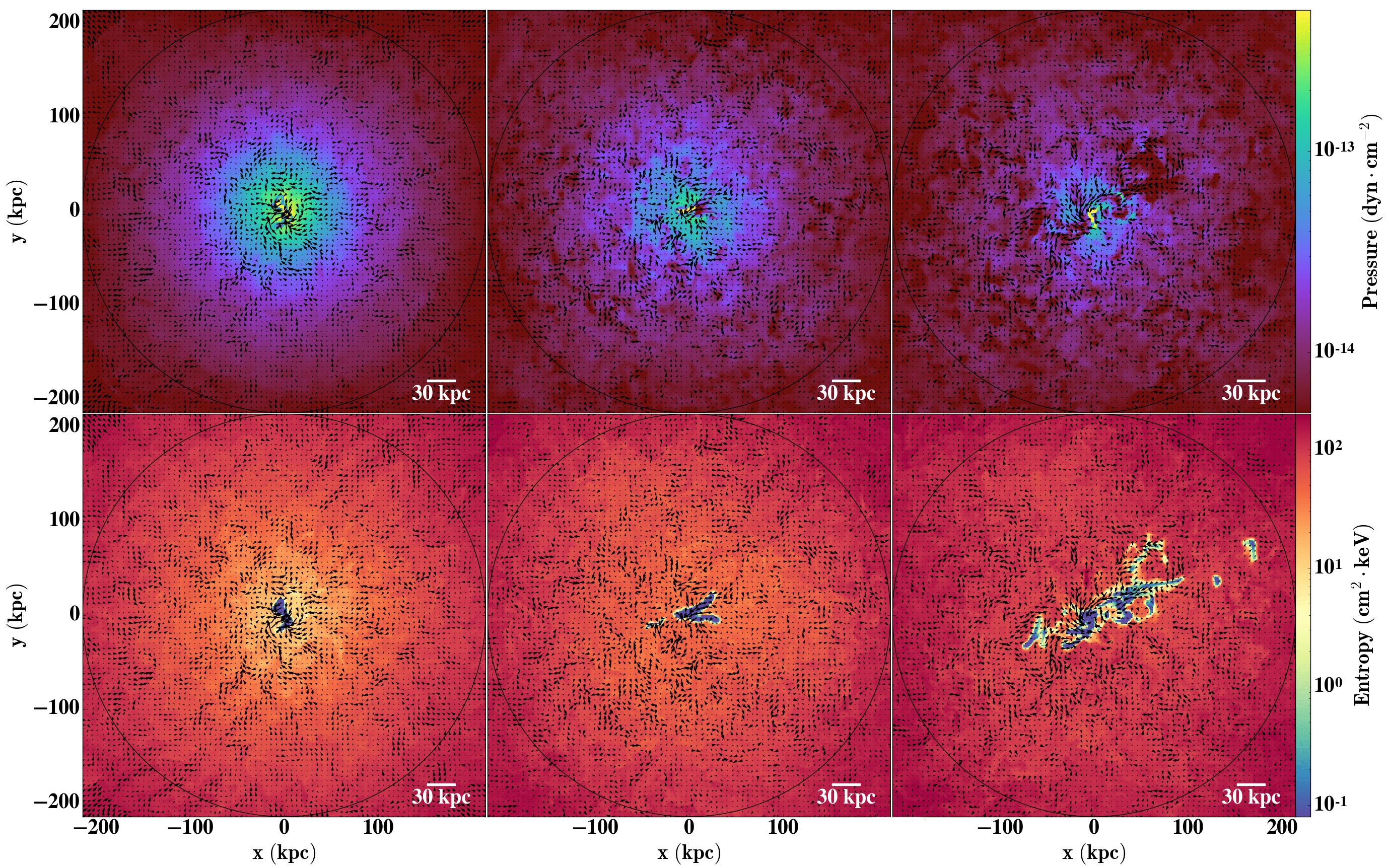}
\caption{Slices at 3 Gyrs along the z-axis showing the Pressure (top row) and Entropy (bottom row) for the Low (left), Mid (middle), and High (right) runs. The arrows show the $x-y$ velocity vectors of the flow and a black circle shows the virial radius at $R_{\rm vir} \approx 220$ kpc. Note that the cooling streams seen in the Mid and High runs are much larger than the driving scale of the turbulence, and they arise as a result of the convective flow that is set up by the turbulent heating.}
\label{fig:pressure_3Gyr}
\end{figure*}

\begin{figure*}[t]
\centering
\includegraphics[width=1.0\linewidth]{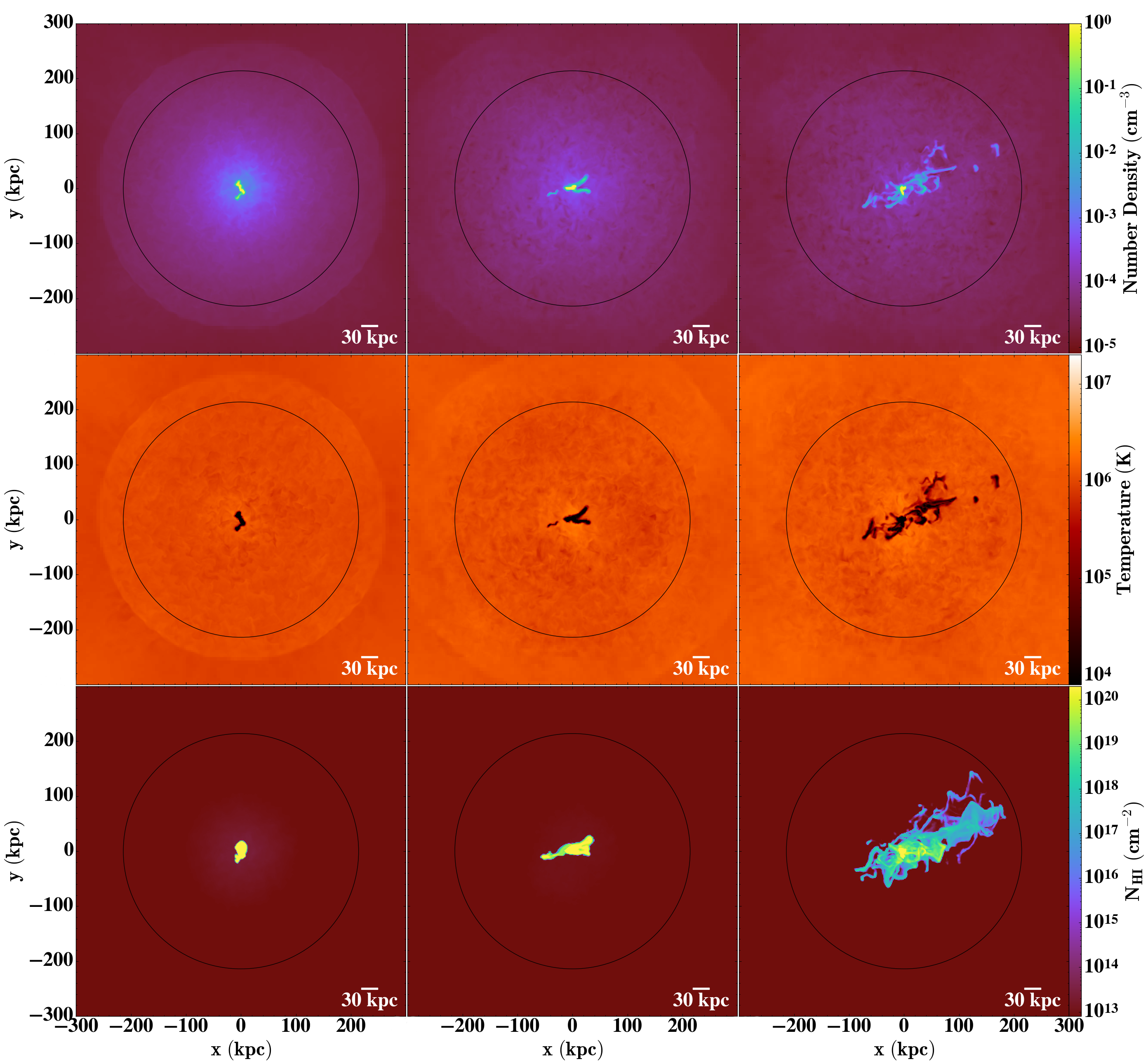}
\caption{Slices at 3 Gyrs along the z-axis showing the number density (top row), temperature (middle row) along with $N_{\rm H\ I}$ projections (bottom row) for the Low (left), Mid (middle), and High (right) runs. A black circle shows the virial radius at $R_{\rm vir} \approx 220$ kpc.}
\label{fig:ndens_3Gyr}
\end{figure*}
\subsection{Model Parameters} \label{model_params}

We conducted a suite of simulations that assumed a Milky Way mass dark matter halo of $M=10^{12}$~M$_{\odot}$, with a virial radius of 220~kpc and a concentration parameter of 10. The medium was initialized with a fractional ion abundance at 0.3~Z$_{\odot}$ metallicity in collisional ionization equilibrium (CIE) at the virial temperature of the halo, $T = 1.2 \times 10^{6}$~K and average sound speed of 166~km~s$^{-1}$.  In all cases the material was assumed to be irradiated with a redshift zero \citet{2012ApJ...746..125H} (HM2012) EUVB whose specific intensity was normalized to $8.23 \times 10^{-24}$ erg cm$^{-2}$ s$^{-1}$ Hz$^{-1}$ sr$^{-1}$ at the Lyman limit following the radio-quiet AGN spectral index ($\alpha=0.157$).  Note that our simulations assume chemical homogeneity, while observations indicate that the composition of the CGM is likely to be significantly inhomogeneous \citep{Zahedy2019}.  However, the impact of this additional factor is left for future investigation.

The simulations were carried out in an 800~kpc box with periodic boundaries using Static Mesh Refinement (SMR) to accurately capture important structures that develop in the halo as turbulence ensues. Refinement begins at $\approx$ 300~kpc and continues into the center. The levels of refinement were as follows: for R $\gtrsim$ 300~kpc the domain was at a resolution of 64$^{3}$ which translates to 12.5~kpc, for 300 $\gtrsim$  R $\gtrsim$ 250~kpc the resolution was 128$^3$ which translates to 6.2~kpc, 250 $\gtrsim$  R $\gtrsim$ 225~kpc the resolution was 256$^3$ which translates to 3.1~kpc, and for 225~kpc $\gtrsim$ R the resolution was 512$^{3}$ which translates to 1.6~kpc. 

The turbulence had stirring modes with wavenumbers between $L/3 \leqslant 2\pi/k \leqslant L$, where we choose $L$ to be 30~kpc. This choice in driving scale is informed by the size of the Milky Way disk as we imagine the turbulence to be driven by outflowing and inflowing processes centered around the disk of the host galaxy. Each of our simulations was run for 3~Gyrs until it reached a {global equilibrium such that the change in total energy was less than 0.25\%}. 

We looked at 3 cases for the average $\sigma_{1D}$: 20, 34, and 41~km~s$^{-1}$ which we call the Low, Mid, and High runs respectively. These follow estimates of the non-thermal velocity components of line-width measurements from W16. In Figure \ref{fig:velocity} we show the volume-averaged $\sigma_{\rm 1D}$ found by subtracting the infall velocity from the volume-averaged velocity within the virial radius of the halo along with their respective mach numbers computed as the ratio of $\sigma_{\rm 1D}$ and the volume-averaged sound speed at 0~Gyr. 

\section{Results} \label{results}

\begin{figure}
    \centering
    \includegraphics[width=1.0\linewidth]{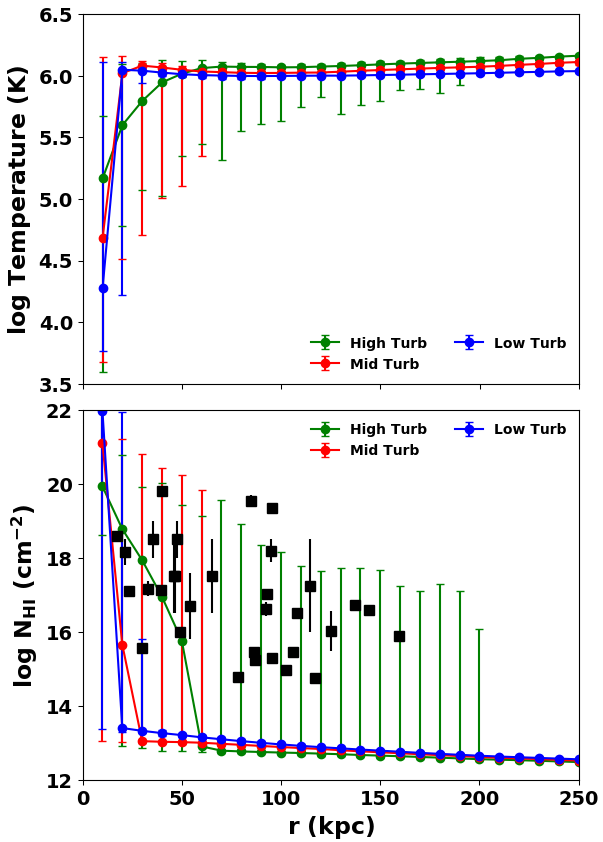}
\caption{{Filled circles show the 50$^{\rm th}$ percentile of sightlines in 10~kpc bins} for the log~Temperature (top) and $N_{\rm H\ I}$ (bottom) vs. radial distance $r$ for the Low (blue), Mid (red), and High (green) runs at 3~Gyrs contained to cells within the virial radius. {Error bars show the 1$^{\rm st}$ percentile as the lower limit and 99$^{\rm th}$ percentile as the upper limit.} Projections were generated for each sightline as a line integral of the ion number density along the $z$-axis. The $N_{\rm H\ I}$ data from the COS-Halos sample overlaid are on our projected columns in black.}
    \label{fig:temp_H1}
\end{figure}

\subsection{Evolution} \label{evolution}

Initially, gas within $r \approx 30$~kpc cools over $\approx 100$~Myrs, forming a low-pressure region in the center of the halo, which promotes an accretion flow. This results in an accretion shock that forms near the core and gradually moves outward. As turbulence develops, it starts to disrupt the flow, such that the cool gas  near the center is anisotropic and filamentary.  Turbulence also acts to stochastically heat the medium, creating buoyant patches of higher entropy gas that moves outward to replace the cooling inflowing gas. This sets up a convective flow that is most extended in the run with the highest level of turbulence. 

This flow may be observed in Figure~\ref{fig:pressure_3Gyr} which shows slices of the pressure and entropy at 3~Gyrs overlaid with black quivers that show the velocity field. Here we see that gas can flow inwards through the low pressure, low entropy structures into the center {where it may cycle out of the center or simply collect there. These low pressure, low entropy structures} house lower ionization state gas as we discuss in more detail below. 

Note also that the Low run displays a profile that is close to spherically symmetric, while the Mid run develops low pressure/entropy systems near the center of the halo. Finally, the High run shows an extended convection flow that can naturally facilitate the development of low pressure/entropy systems at and beyond $100$~kpc, forming low ionization state gas at large distances from the halo center. 

This figure also illustrates how the overall entropy profile varies with increased levels of turbulence. In the Low run, the entropy increases strongly with radius, indicating a medium that is largely stable to convection. In the Mid run, on the other hand, the entropy is much more constant near the center, which hosts a limited convective region. Finally, the High run has the largest convective regions and {a more gradual radial increase in entropy.  This profile is the most similar to the CGM produced in the high mass loading star formation (SF) feedback model in \citet[][hereafter F17]{fielding2017impact} as well as the model in \citet{li2019supernovae} with a star formation rate of 3~M$_{\odot}$~yr$^{-1}$ and the turbulent core simulations in \citet{su2020cosmic}}. In the High run, higher entropy gas is found throughout the halo, intermixed with low-entropy gas. In between these two are layers of intermediate entropy gas ($S \approx 10$~keV~cm$^{-2}$), which cools rapidly along the interfaces between the cold and hot material. 

Overall we find all of the species in the simulations vary by $\lesssim$ 5\% for the last 100~Myrs of the runtime. The elements that are most out of equilibrium differ across the runs, however, we find Ne, Ca, and S to be common elements with the highest percentage difference from equilibrium. Of these, Ne is the least surprising, as it has the most ions that we track in our reaction network. 

\subsection{General Features} \label{gen_features}

To give a more complete picture of the structure of the gas at 3~Gyrs, in Figure~\ref{fig:ndens_3Gyr} we show slices of the number density and temperature, along with projections of $N_{\rm H\ I}$ for each of our runs. Here as in Figure~\ref{fig:pressure_3Gyr}, the Low run shows a nearly spherically symmetric distribution. Furthermore the temperature distribution within the accretion shock is largely homogeneous, while density falls off strongly with radius. Meanwhile, the Mid and High cases, with their increased levels of turbulence, produce more structures within the halo. The High run, in particular, shows small clumps of cold ($T \approx 10^{4}$~K) gas at nearly the virial radius of the halo, and a density profile that falls off more slowly with radius. We also see large amounts of \ion{H}{1} constrained to the inner 30 and 60~kpc of the Low and Mid runs respectively, while the High run is turbulent enough to facilitate cool, inflowing \ion{H}{1} out to 200 kpc, as seen in the positive x/y quadrant.

\begin{figure*}
\centering
\includegraphics[width=1.00\linewidth, height=1.0\textheight]{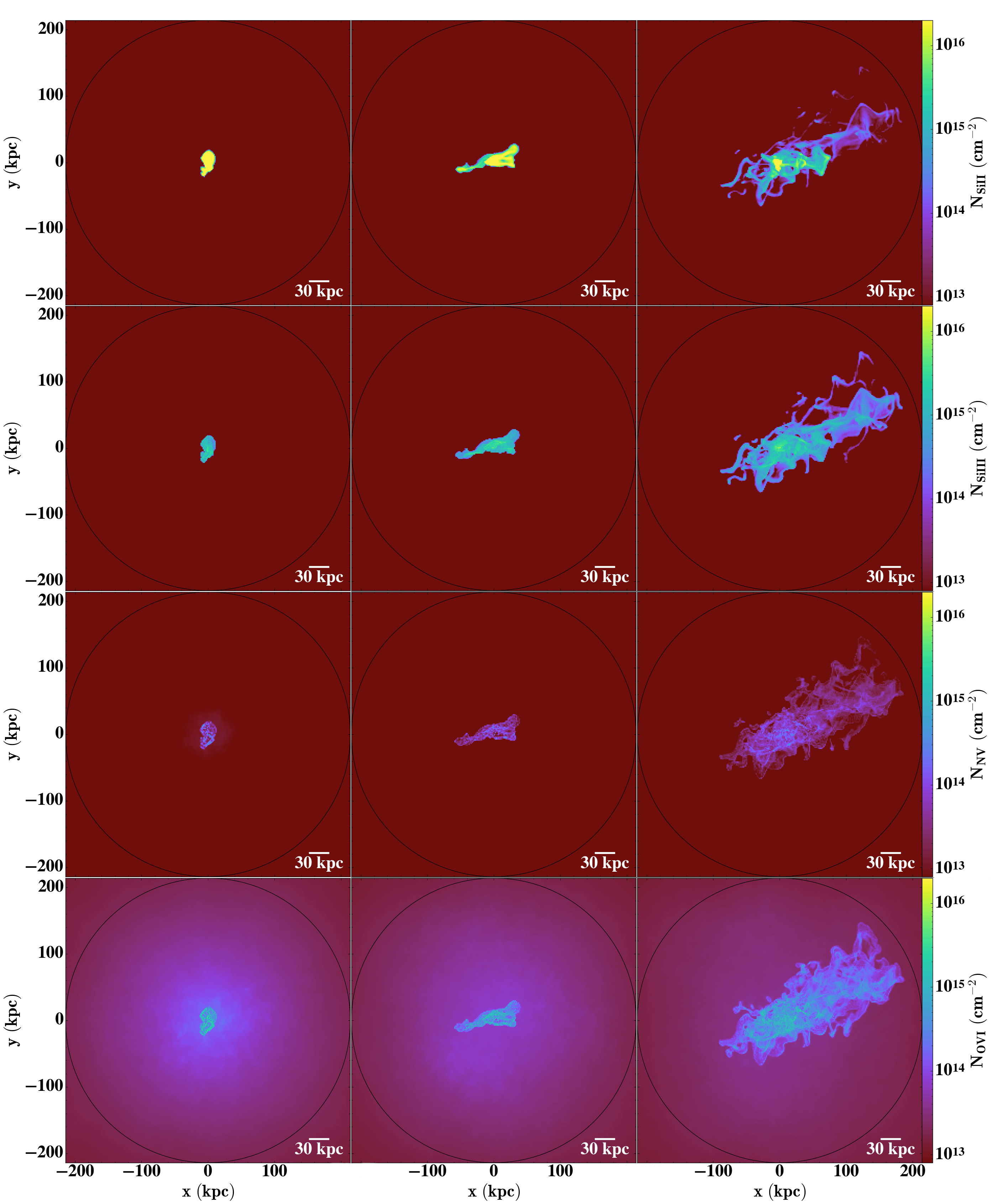}
\caption{$N_{\rm Si\ II}$ (first row), $N_{\rm Si\ III}$ (second row), $N_{\rm N\ V}$ (third row), and $N_{\rm O\ VI}$ (fourth row) projections for the Low (left column), Mid (middle column), and High (right column) runs. In all panels, the black ring shows the virial radius at $r \approx 220$~kpc.}
\label{fig:all_ions_profiles}
\end{figure*}

In Figure \ref{fig:temp_H1}, we show the density-weighted temperature and $N_{\rm H\ I}$ {taken as the 50$^{\rm th}$ percentile of the spread in 10~kpc bins} vs. radial distance $r$, {where projections are} integrated along the $z$-axis, as compared to observations from the COS-Halos sample. The median density-weighted temperature does not vary significantly from the equilibrium temperature of the halos, however, the increased levels of turbulence {yield a wider spread in the temperature at higher radii and throughout the halo as} seen in the temperature slices. 

As a consequence, the $N_{\rm H\ I}$ distribution shows a steep decrease at $\approx$ 20~kpc for the Low {and Mid} runs which becomes more gradual in the High run.
 The results from this run match well with the COS-Halos data, except for 2 systems which lie above the 99$^{\rm th}$ percentile limits. The High run also shows a high covering fraction of Lyman Limit Systems (defined as systems with $N(HI) > 10^{17}$ cm$^{-2}$) out to $\approx 50$ kpc of the virial radius.  This is roughly consistent with the observations of \cite{Chen2018}, who found a high covering fraction of such systems out to $\approx 150$ kpc in a sample of luminous red galaxies, likely associated with massive halos with virial radii $2-3$ times larger than the ones considered here \citep[e.g.][]{Artale2018, McEwen2018, Zehavi2018}.  The N$_{\rm H\ I}$ trend from the High run also matches high mass loading stellar feedback model in F17, as well as the observations of the extended CGM, described in \citet{2015johnson}.

Note that when looking at the projections of \ion{H}{1} we can say that the sightlines that match the COS-Halos data after about 100~kpc are from the inflowing material in the positive x/y quadrant and not uniformly distributed around the halo.  This may or may not  be representative of all star-forming galaxy halos, as Lyman~$\alpha$ emission has been observed at these impact parameters in $z \approx 2.5$ galaxies \citep{cantalupo2014cosmic,prescott2015spatially}.

\subsection{COS-Halos Comparisons} \label{cos_halos_compare}

\begin{figure*}[t]
    \centering
    \includegraphics[width=1.0\linewidth]{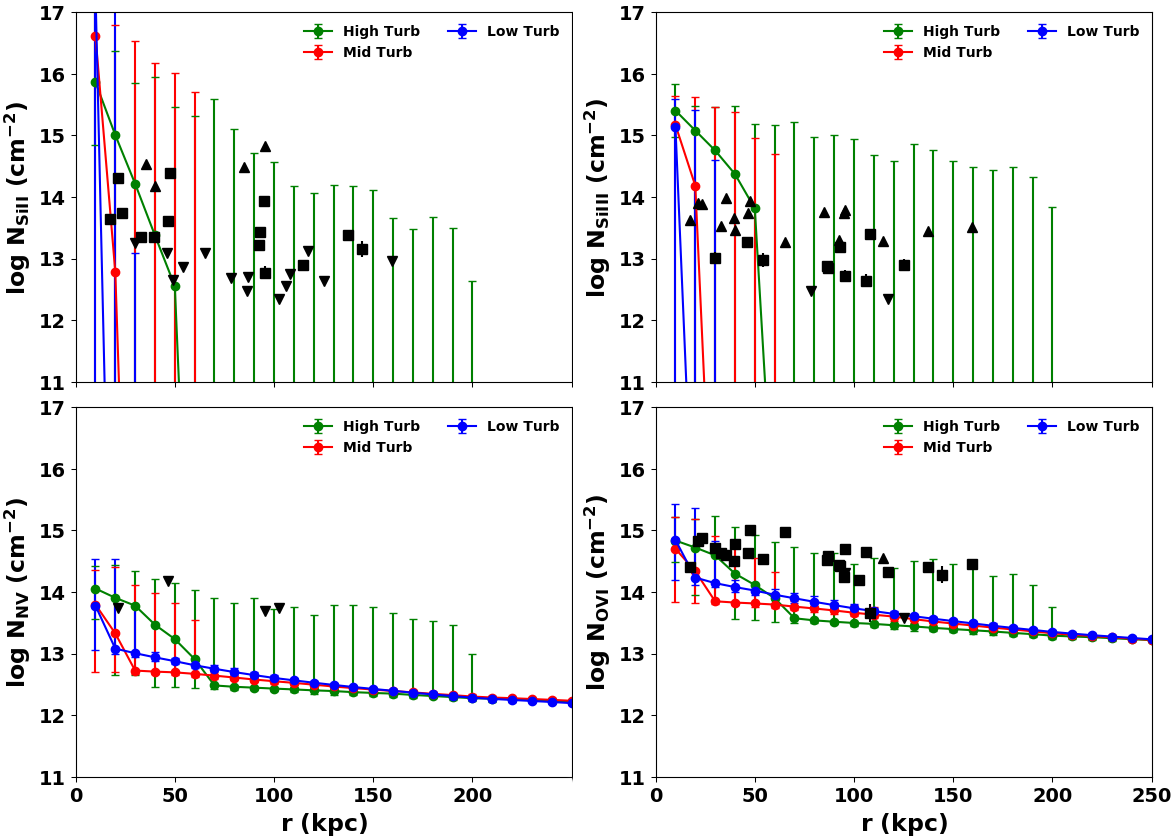}
    \caption{{Filled circles show the 50$^{\rm th}$ percentile of sightlines in 10~kpc bins} for Log $N_{\rm Si\ II}$ (top left), $N_{\rm Si\ III}$ (top right), $N_{\rm N\ V}$ (bottom left), and $N_{\rm O\ VI}$ (bottom right) vs. radial distance $r$ for the Low (blue), Mid (red), and High (green) runs at 3~Gyrs contained to cells within the virial radius. {Error bars show the 1$^{\rm st}$ percentile as the lower limit and 99$^{\rm th}$ percentile as the upper limit.} Projections were generated for each sightline as a line integral of the ion number density along the $z$-axis. We also show detections (black squares) and limits (black arrows) from the COS-Halos sample overlaid on our projected column densities.}
    \label{fig:all_ions}
\end{figure*}

In Figure~\ref{fig:all_ions_profiles}, we show the projected column density maps for several key ions, and in Figure~\ref{fig:all_ions} we plot the
corresponding column density profiles, {again showing the 50$^{\rm th}$ percentile of sightlines vs. radial distance $r$ in 10~kpc bins}. In the top two rows of Figure~\ref{fig:all_ions_profiles}, we focus on  \ion{Si}{2} and \ion{Si}{3}, which sample low and intermediate ionization state material commonly observed in the halos of low-redshift galaxies \citep{2013werkApJS..204...17W,2016borthakurApJ...833..259B,2017heckman}. In the bottom two rows, we show \ion{N}{5}, which has an ionization potential just 40~eV below the 138 eV ionization potential  of \ion{O}{6}, and is chosen due to its lack of absorption in the CGM of star-forming galaxies for reasons that remain to be determined (W16), and \ion{O}{6}, an ion known to trace the hotter ambient gas within the halo \citep{sembach2003highly,werk2016ApJ...833...54W,2017reviewARA&A..55..389T}.

This figure shows that $N_{\rm Si\ II}$ is mostly concentrated in the center as this contains moderately dense ($n \approx 10^{-3}$~cm$^{-3}$), cold ($T \approx 10^{4}$~K) clumps. The Low and Mid runs show this low ionization state material being tightly constrained to the inner 50~kpc, while the High run shows \ion{Si}{2} extending out to near the virial radius of the halo, embedded within the large inflow of material.

The column density profiles for $N_{\rm Si\ II}$ at low impact parameter show agreement between all of the runs and the limits/detections from the COS-Halos sample. {However, after about 20~kpc the Low and Mid runs show steep declines in their profiles, that do not agree with the observed detections (note COS-Halos limits and detections lie within the 99$^{\rm th}$ percentile upper limits of the Mid run up to $\approx$ 60~kpc). However, a more gradual decline in $N_{\rm Si\ II}$ upper limits is seen in the High run, which agrees well with the COS-Halos results.}

The intermediate ionization state component, as traced by \ion{Si}{3}, also shows a fairly different view across the varying levels of turbulence in the second row of the column density maps plotted in Figure~\ref{fig:all_ions}. \ion{Si}{3} follows a similar spatial distribution to \ion{Si}{2}, but traces slightly hotter material surrounding the coldest gas. The $N_{\rm Si\ III}$ profiles also show a similar picture to $N_{\rm Si\ II}$ with an agreement between all of the runs and the COS-Halos sightlines, {followed by a steep decline at about 20~kpc for the Low and Mid runs while the more steady decline in upper limits of the High run agrees well with the observed detections and limits.}

In Figures \ref{fig:project1}, \ref{fig:project2}, and \ref{fig:highres_columns} in Appendix A we show projected column density maps for the other COS-Halos ions (\ion{C}{2}, \ion{C}{3}, \ion{Mg}{2}, \ion{N}{2}, \ion{O}{1}, and \ion{Si}{4}, omitting the iron ions) as well as COS-Halos observations. These column density maps show similar trends as the ones above, and the radial trends in ionic columns from the High run again shows a good agreement with the COS-Halos data. The trends also match well with the recent $z \lesssim 1$ COS CGM Compendium survey described in \citet{lehner2018cos}, and other studies of the cool CGM around the nearby galaxies M31 in \citet{lehner2015evidence} and NGC~1097 in \citet{bowen2016structure}. They also are in reasonable agreement with fully cosmological simulations such as the Evolution and Assembly of Galaxies and their Environments \citep[EAGLE;][]{oppenheimer2018multiphase}, Feedback in Realistic Environments \citep[FIRE;][]{ji2019properties}, and IllustrisTNG \citep[][]{Kauffmann2019}.

Next we look at \ion{N}{5} and \ion{O}{6} as shown in the bottom 2 rows of Figure~\ref{fig:all_ions}. The \ion{N}{5} within the halo shows similarities with the \ion{O}{6}, however, it has a significantly lower abundance, a feature more evident in the column density profiles in Figure~\ref{fig:all_ions_profiles}. $N_{\rm O\ VI}$ {limits/detections lie within the 1$^{\rm st}$ to 99$^{\rm th}$ percentile errorbars of the Low and Mid runs at for r < 60~kpc} while limits/detections further out agree well with the High run. The \ion{O}{6} columns observed in the extended CGM study of \citet{2015johnson} show gradually decreasing columns that lie within our 1$^{\rm st}$ to 99$^{\rm th}$ percentile errorbars. {The radial decrease in abundance of both of these ions in Figure~\ref{fig:all_ions} and their distribution in Figure~\ref{fig:all_ions_profiles}, lead us to theorize that sightlines in which \ion{O}{6} is observed without accompanying \ion{N}{5} may primarily trace ambient halo gas, while sightlines that show both \ion{O}{6} and \ion{N}{5} absorption might instead trace material in a shell around cooler inflowing gas.}

\begin{figure*}
\centering
\includegraphics[width=1.0\linewidth]{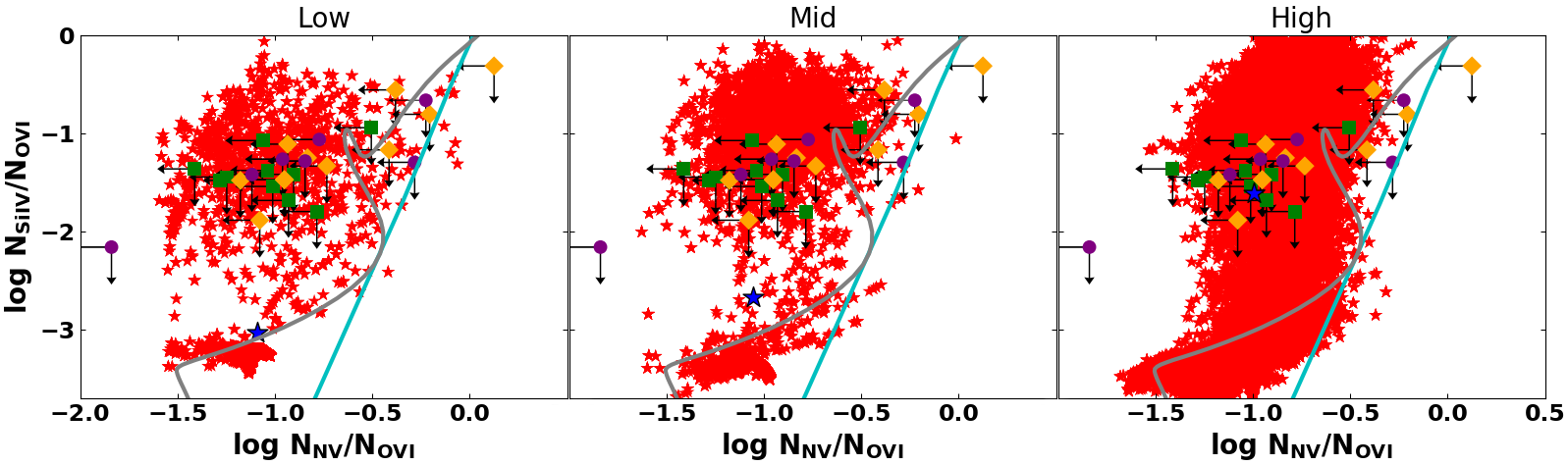}
\caption{Log~$N_{\rm Si\ IV}/N_{\rm O\ VI}$ vs. log~$N_{\rm N\ V}/N_{\rm O\ VI}$ from the Low (left), Mid (middle), and High (right) runs are shown as red stars with the averages shown as blue stars. These are overlaid by the matched \ion{Si}{4}, \ion{N}{5}, and \ion{O}{6} components from W16 Figure 12. Green circles represent the ``broad'' type \ion{O}{6} absorption, orange circles are ``narrow'' type absorbers, and purple circles are ``no-low'' type absorbers. As a reference, we plot a single-zone Cloudy model photoionized by an HM2012 EUVB (solid cyan) with -4 < log~$U$ < -1 along with an Orly \& Gnat {time-dependent radiatively cooling} isochoric solar metallicity model in gray.}
\label{fig:Si4_N5}
\end{figure*}

\begin{figure*}
\centering
\includegraphics[width=1.0\linewidth]{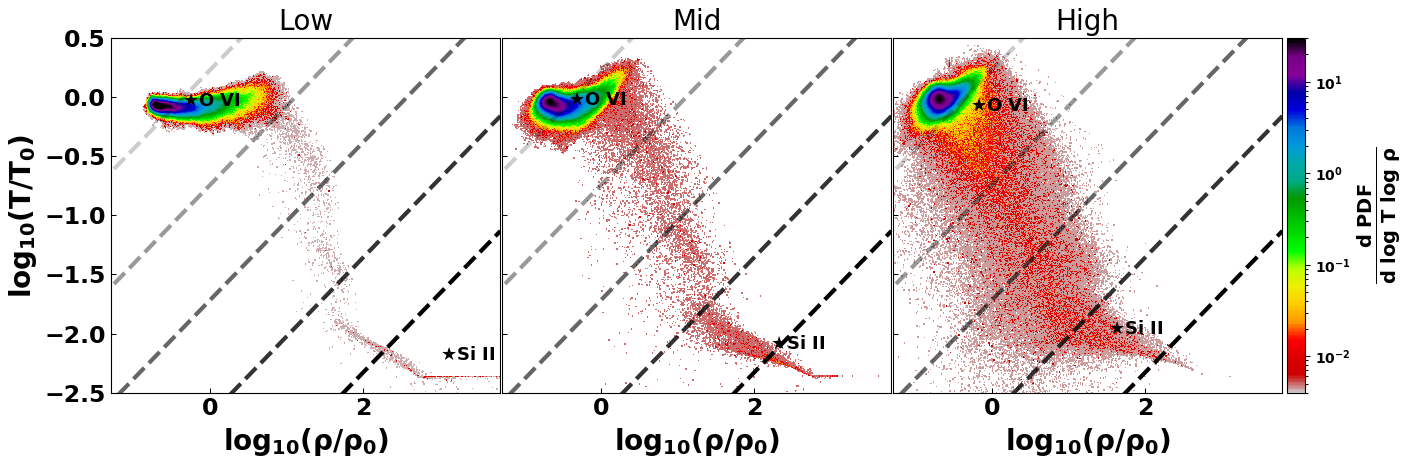}
\caption{Log~$T/T_{0}$ vs. log~$\rho/\rho_{0}$ probability density maps from the Low (left), Mid (middle), and High (right) runs where $T_{0} = 1.2 \times 10^{6}$~K and $\rho_{0} = 2.1 \times 10^{-28}$~g~cm$^{-3}$ are at $t$ = 0. The maps omit data within $r$ = 12~kpc and are overlaid with the \ion{O}{6} and \ion{Si}{2} weighted temperatures/densities as black stars in addition to dashed lines indicating lower (higher) entropy as darker (lighter) lines. The lines of constant entropy are as follows : 0.1, 0.6, 3.1, 17.2, and 95.5~KeV~cm$^{-2}$.}
\label{fig:PDF}
\end{figure*}

{We furthermore compare our \ion{O}{6} with other recent theoretical results, and again find very similar trends between the $N_{\rm OVI}$ profile from our High run and the high mass loading model in F17. The simulations done in that work used equilibrium heating and cooling rates from \citet{wiersma2009effect} that include the effects of a EUVB, which are important at the low densities typical to the CGM. We also find our $N_{\rm OVI}$ profiles to be greater than those given from the chemical equilibrium cooling flow solutions of 11.5 < log~$M_{\odot}$ < 12.5 halos found in \citet{stern2019cooling}, however, the authors state that this may be alleviated with some mechanism to heat the outer CGM or if \ion{O}{6} traces photoionized gas beyond an accretion shock. As we discuss in section \ref{evolution}, the gas in our simulations is shock heated throughout the first phase of its evolution and thus the ambient \ion{O}{6} primarily traces post-shock, photoheated gas, while higher abundances of this ion surrounding cooling flows may primarily be radiatively cooling gas.}

We also compare our results to the $N_{\rm Si\ IV}/N_{\rm O\ VI}$ and $N_{\rm N\ V}/N_{\rm O\ VI}$ ratios observed in \citet{werk2016ApJ...833...54W} in Figure~\ref{fig:Si4_N5} where we only show sightlines with log~$N_{\rm O\ VI}$ > 13.5 to limit our sample to similar sightlines as the COS-Halos sample in W16. Furthermore, we show a single-zone Cloudy model photoionized by an HM2012 EUVB model in cyan and a \citet{2007gnatApJS..168..213G} isochoric solar metallicity model overlaid in grey. 

The Cloudy models have the largest values of $N_{\rm N\ V}/N_{\rm O\ VI}$ and are the most discrepant with the observations, indicating that the large observed differences between  \ion{N}{5} and \ion{O}{6} can not be explained by equilibrium models. Furthermore, \citet{buie2020interpreting} showed that such equilibrium models tend to underpredict the amount of \ion{O}{6} which may explain the high $N_{\rm N\ V}/N_{\rm O\ VI}$ values produced.mThe non-equilibrium isochoric cooling models from \citet{2007gnatApJS..168..213G} lie closer to the data, but still have $N_{\rm N\ V}/N_{\rm O\ VI}$ {values} larger than {most of the} observations, suggesting that radiative cooling alone is insufficient to explain these results. 

Finally, the MAIHEM results, which include not only radiative cooling but cooling by the mixing between cold and hot material, have a shape that is similar to the \citet{2007gnatApJS..168..213G} results, but shifted further to the left, covering the observed range of $N_{\rm N\ V}/N_{\rm O\ VI}$ ratios. This suppressed ratio arises from gas that cools even faster than isochorically, due the combination of radiative cooling and the turbulent mixing of hot and cold media. This ratio is similar for all three runs with the log of the ratios being -1.0, -1.05, and -1.09 for the Low, Mid, and High runs respectively.

The similar dispersion in our model ratios suggests that while strong turbulence is needed to set up a large-scale convective flow in the halo, leading to cold gas at large radii, it is not essential to reproducing the $N_{\rm N\ V}/N_{\rm O\ VI}$ results. Rather it is a feature that is common to all of our runs: the rapid cooling of gas through simultaneous mixing and radiation, which results in the extreme non-equilibrium ratios observed in nature. Note also that this ratio is not a result of the log~$N_{\rm O\ VI}$ > 13.5 column density limit imposed by the observations. If we relax this limit we simply see a greater amount of sightlines that agree with the observed $N_{\rm N\ V}/N_{\rm O\ VI}$ values. 

{Our average $N_{\rm N\ V}/N_{\rm O\ VI}$ is also consistent with ratios from recent re-observations of two CGM systems in \citet{lochhaas2019high} and some of the upper limits implied by these observations, however, we do not have any sightlines that fall below log~$N_{\rm N\ V}/N_{\rm O\ VI}$=-1.8, implying some other physics may be necessary to fully capture the dynamic CGM, such as thermal conduction and/or magnetic fields. \citet{liang2020model} showed in their simulations of cold clouds in a hot medium that these considerations increase the efficiency of heat transfer from cooling gas while a weak magnetic field acts to shield developing cold clouds. This would reduce the lifetime of intermediate ions tracing gas transitioning between high and low ionization states while maintaining small clumps of the low ionization state material.}

{Finally, the average $N_{\rm Si\ IV}/N_{\rm O\ VI}$ values in the MAIHEM models, increase with greater turbulence. This is a result of the increased amount of colder gas in the CGM of the more turbulent, runs which encompasses larger amounts of \ion{Si}{4}. When only considering gas with $n \leq$ 10$^{-4}$~cm$^{-3}$, the resulting log~$N_{\rm Si\ IV}/N_{\rm O\ VI}$ values are $\lesssim$ -0.5. }

Finally, we plot the 2D, volume-weighted probability density distribution for our runs in Figure~\ref{fig:PDF}, which indicates the phase distribution of the media. Here we observe that most of the volume lies at a value slightly lower density than the average at $t$ = 0~Gyr and virial temperature of the halo. Increasing turbulence not only disperse a greater number of cells across lower temperatures and densities but also promote a trend where the majority of cells go towards a constant entropy, as expected for a convective flow. Finally, increasing turbulence also pushes the \ion{Si}{2} weighted temperature/density towards less dense and hotter gas while the \ion{O}{6} weighted temperature/density varies little between the runs.

\begin{figure*}
\centering
\includegraphics[width=1.0\linewidth]{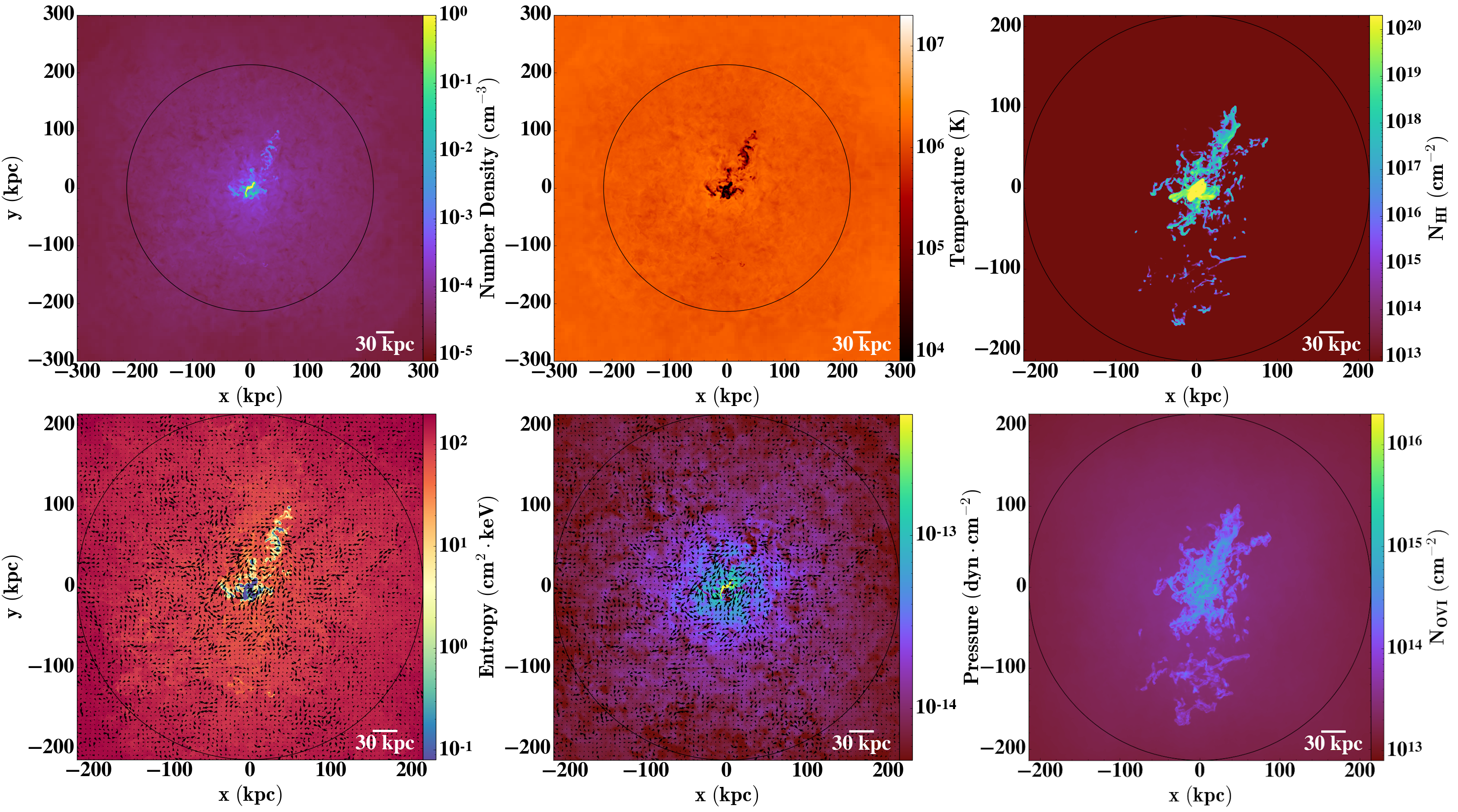}
\caption{Panels showing the number density slice (top left), temperature slice (top middle), projected \ion{H}{1} (top right), entropy slice (bottom left), pressure slice (bottom middle), and projected \ion{O}{6} (bottom right) at 3~Gyrs along the z-axis for the double-resolution High run. The arrows show the $x-y$ velocity vectors of the flow and a black circle shows the virial radius at $R_{\rm vir} \approx 220$ kpc.}
\label{fig:high_res_panel1}
\end{figure*}

\begin{figure}
\centering
\includegraphics[width=1.0\linewidth]{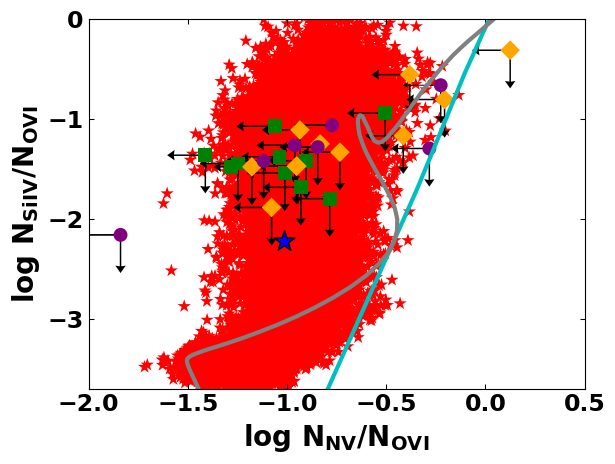}
\caption{Log~$N_{\rm Si\ IV}/N_{\rm O\ VI}$ vs. log~$N_{\rm N\ V}/N_{\rm O\ VI}$ from the double-resolution High run shown as red stars with the average shown as a blue star. These are overlaid by the matched \ion{Si}{4}, \ion{N}{5}, and \ion{O}{6} components from W16 Figure 12. Green circles represent the ``broad'' type \ion{O}{6} absorption, orange circles are ``narrow'' type absorbers, and purple circles are ``no-low'' type absorbers. As a reference, we plot a single-zone Cloudy model photoionized by an HM2012 EUVB (solid cyan) with -4 < log~$U$ < -1 along with an Orly \& Gnat {time-dependent radiatively cooling} isochoric solar metallicity model in gray.}
\label{fig:Si4_N5_highres}
\end{figure}

\subsection{Convergence Tests} \label{converge}

Finally, as a test of convergence, we repeated the High run with double the resolution to see how this affects the results and evolution of the simulation. Specifically, the driving scale and strength of the turbulence were taken to be the same as the High run described above, but the levels of refinement were as follows: for R $\gtrsim$ 300~kpc the domain was at a resolution of 128$^{3}$ which translates to 6.2~kpc, for 300 $\gtrsim$  R $\gtrsim$ 250~kpc the resolution was 256$^3$ which translates to 3.1~kpc, 250 $\gtrsim$  R $\gtrsim$ 225~kpc the resolution was 512$^3$ which translates to 1.6~kpc, and for 225~kpc $\gtrsim$ R the resolution was 1024$^{3}$ which translates to 0.8~kpc. 

\begin{figure}
\includegraphics[width=1.00\linewidth]{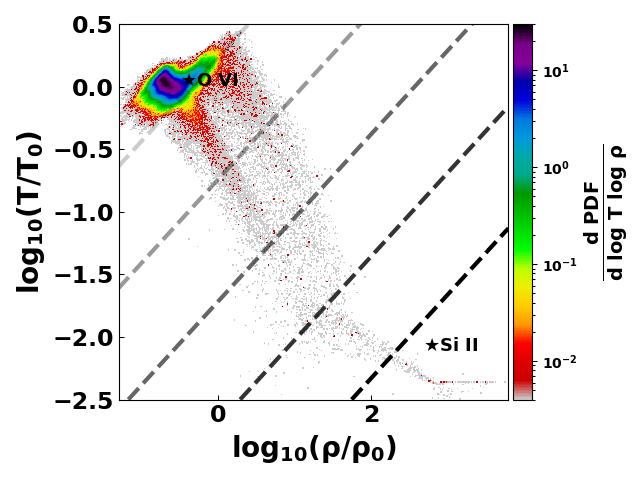}
\caption{Log~$T/T_{0}$ vs. log~$\rho/\rho_{0}$ probability density maps from the double-resolution High run where $T_{0} = 1.2 \times 10^{6}$~K and $\rho_{0} = 2.1 \times 10^{-28}$~g~cm$^{-3}$ are at $t$ = 0. The maps omit data within $r$ = 12~kpc and are overlaid with the \ion{O}{6} and \ion{Si}{2} weighted temperatures/densities as black stars in addition to dashed lines indicating lower (higher) entropy as darker (lighter) lines. The lines of constant entropy are as follows : 0.1, 0.6, 3.1, 17.2, and 95.5~KeV~cm$^{-2}$.}
\label{fig:highres_pdf}
\end{figure}

This run evolves similarly to the previous runs, with gas within a 30~kpc radius cooling rapidly for the first 100~Myrs followed by an outward moving accretion shock at 1~Gyr. The average 1D velocity dispersion of this run was 40~km~s$^{-1}$, extremely similar to the 41~km~s$^{-1}$ we found in the lower resolution case with the same driving. 

However, due to the increased resolution, there is more cooling in the initial stages, which {eventually settles to similar cooling rates as compared to its original resolution counterpart. A consequence of the increased cooling during the initial stages is that} more material {is able} to collapse towards the center through the accretion shock. This results in the density and temperature profiles shown in Figure~\ref{fig:high_res_panel1}, along with pressure and entropy profiles {that are} similar to a mix between the Mid and High runs. In this double-resolution case we still see the formation of accretion flows similar to those found in the original High run that feeds gas from the hot ambient halo through low pressure/entropy filaments. The higher resolution also allows for smaller \ion{H}{1} clouds to exist further out in the CGM in addition to filaments. 

The increased resolution leads to the formation of smaller structures as compared to the {larger} filamentary inflows observed in its original resolution counterpart, as visible in the column density maps of \ion{H}{1}  and \ion{O}{6} shown in Figure~\ref{fig:high_res_panel1}, as well as a wide range of ions as shown in Figure~\ref{fig:highres_maps} in the appendix.

{Looking at the column density profiles for all our runs in Figure~\ref{fig:highres_columns} in the appendix, we find that the 50$^{\rm th}$ percentiles from this run typically decrease faster than its original resolution counterpart, owing to its more centrally concentrated density profile which is a result of the larger amounts of cooling in the initial stages. The COS-Halos limits and detections remain in agreement with the errorbars of the column density profiles, however, they show smaller columns $\gtrsim 100$~kpc compared to those from the original resolution run due to the more resolved cooling. Higher feedback in the center, which could take the form of higher velocity dispersion, a more negative $\alpha$ in Equation~\ref{equ:trend}, or rotation in the center would likely reduce the amount of gas cooling towards the center while maintaining smaller structures that can cool out of the ambient medium as opposed to the filaments of gas in the original resolution simulation.}

{When these column densities are converted into ratios, we see a very similar trend in $N_{\rm Si\ IV}/N_{\rm O\ VI}$ vs. $N_{\rm N\ V}/N_{\rm O\ VI}$ as shown in Figure~\ref{fig:Si4_N5_highres} as compared to the original resolution run. Like the original High run, the increased resolution run produces many sightlines with ratios that agree with limits found in W16, as well as many that trace higher density gas, which lies above the $N_{\rm Si\ IV}/N_{\rm O\ VI}$ upper limits. }
 
Finally, in Figure~\ref{fig:highres_pdf} we show the 2D probability distribution for the double-resolution High run. Here we see that the core of the distribution is extended along the direction of constant entropy, pointing to a large scale convective flow similar to that in the original resolution High run.  However, the higher resolution results in a double-peaked tail towards lower temperatures/densities that is similar to the Low turbulent run.

\section{Constraints on CGM Turbulent Motions} \label{constraints}

\subsection{Observational Constraints on CGM Turbulence} \label{obs_constraints}

While the levels of turbulence incorporated in our High run provide a good match to multiphase measurements of ions in the low redshift CGM, they are also somewhat larger than those inferred from Doppler $b$ parameters in individual low ionization state absorbers. Recently, \cite{Zahedy2019} compared the Doppler parameters of \ion{Mg}{2} and \ion{H}{1} absorbers within the inner CGM of massive elliptical galaxies at $z \approx 0.4.$  Using ratios of the $b$ values of these ions, they found that the gas associated with low ionization state gas has a mean temperature of $2 \times 10^4$K and a modest non-thermal broadening of $\approx 7$~km~s$^{-1}$. On the other hand, W16 and \cite{Zahedy2019} found fairly broad ($b$ > 40~km~s$^{-1}$) \ion{O}{6} absorption in the CGM of star-forming galaxies with an average non-thermal contribution of $\approx$ 40-50~km~s$^{-1}$ although possible blending of narrow components with velocity offsets remains an issue for these measurements, due to the moderate resolution of the COS spectrograph.

At $z=2,$ the Keck Baryonic Structure Survey (KBSS) used High-Resolution Echelle Spectrometer (HIRES) measurements of 130 metal-bearing absorbers to determine the $b$ parameters in the CGM of $\approx L^*$ galaxies. They found that \ion{O}{6} and \ion{C}{4} absorbing gas resulted from somewhat hotter gas than low-redshift \ion{Mg}{2}, but the non-thermal width of this material was  $\approx 6$~km~s$^{-1}$ \citep{Rudie2019}.

\begin{figure*}
\includegraphics[width=1.00\linewidth]{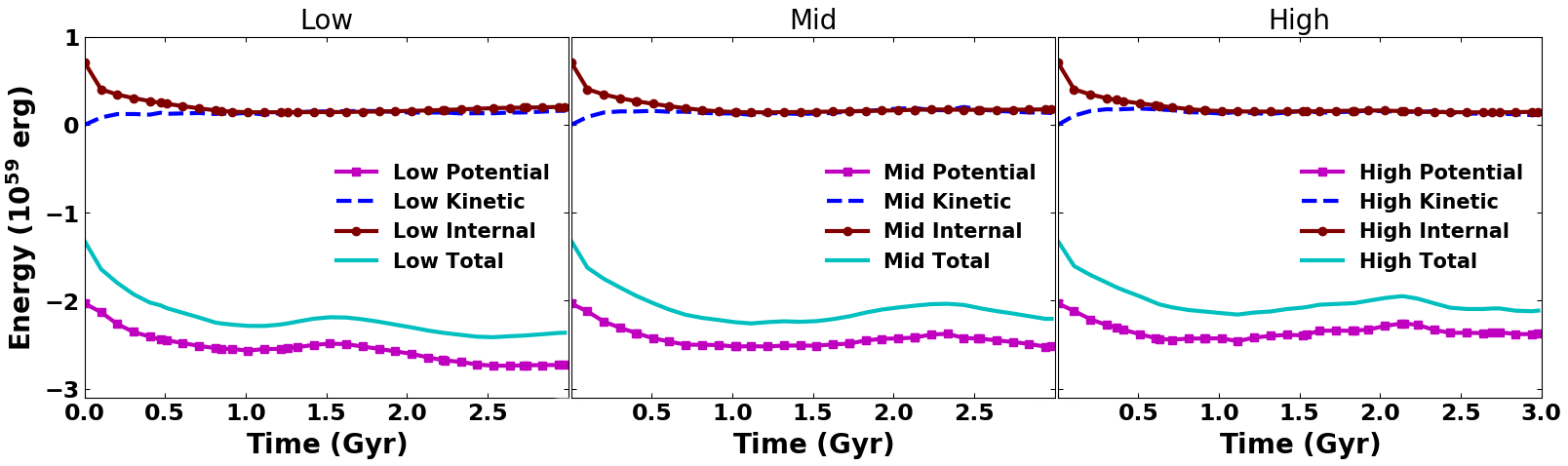}
\caption{Energy vs time in potential (magenta squares), kinetic (blue dashed), internal (maroon circles), and total (solid cyan) for the Low (left), Mid (middle), and High (right) runs. The energy is shown in units of 10$^{59}$~erg.}
\label{fig:energy}
\end{figure*}

It is important to keep in mind, however, that these measurements constrain the turbulent velocities associated with single absorbers, which due to the turbulent cascade, are smaller than the velocities at the driving scale by a factor of $(L_{\rm absorber}/L_{\rm drive})^\alpha$ where alpha is $1/3$ to $1/2$ depending on the Mach number of the medium \cite{Kritsuk2007,pan2010}. The sizes of single cold absorbers in our simulations are near the resolution limit of $(L_{\rm absorber} \approx 2$~kpc).  This means that the expected observed $b$ parameters of cold ions in the high run should be about 10~km~s$^{-1},$ consistent with the observations of $\approx 10^{12}$~M$_\odot$ halos.

The overall level of turbulence is somewhat better constrained in galaxy clusters and groups, where line emission in the intracluster medium allows for measurements at physical scales at or above the driving scale.  Such measurements have been carried out directly in the Perseus Cluster of galaxies by the Hitomi satellite, which found $160-200$~km~s$^{-1}$ motions in the central regions, corresponding to an energy density in turbulent motions of $\approx 4\%$ or a Mach number of $\approx 0.25$ \cite{Hitomi2016,Hitomi2018}. These values can also be compared to those in a much larger range of objects, as measured by a variety of indirect techniques, including resonant scattering, X-ray surface brightness fluctuation analysis, and measurements of the kinematic Sunyaev-Zeldovich effect \citep{Simionescu2019}.

The resonant scattering approach relies on the fact that several of the brightest X-ray emission lines are moderately optically thick, causing them to be suppressed at a level that is highly dependent on turbulent broadening \citep{Gilfanov1987,Churazov2004}.  In clusters, this has led to controversial results \citep{Kaastra1999,Mathews2001,Sanders2006}, due to complications in interpreting the He-like Fe line at 6.7~keV because of variations of gas temperature and metallicity \citep{Zhuravleva2013}. On the other hand, the results have been clearer around massive elliptical galaxies, in which suppression of the Ne-like Fe line at 15.01~\AA\ was detected \citep{Xu2002,Werner2009}. Most recently, \cite{Ogorzalek2017} obtained velocity constraints for 13 such massive galaxies, determining an averaged 1D Mach number of 0.25, corresponding to a non-thermal energy fraction of $\approx 10\%$. 
 
A second approach is to relate the observations of X-ray surface brightness and gas density fluctuations to the velocity fluctuations in the underlying medium \citep{Gaspari2013}. In this way, \cite{Hofmann2016} analyzed deep Chandra observations of 33 well-known clusters, arguing that the observed fluctuations correspond to a sample averaged 1D Mach number of $0.16$. Similarly \cite{Zhuravleva2018} conducted a statistical analysis of X-ray surface brightness and gas density fluctuations in the cool cores of 10 nearby clusters concluding the non-thermal energy was $\approx 5\%$ of the thermal energy in the inner half-cool-core regions and up to $12\%$ in the outer core regions. 

In massive clusters, a third method of estimating the integrated contribution of the non-thermal pressure relies on the fact that measurements of the total baryon fraction of massive clusters are largely insensitive to uncertainties in baryonic physics such as cooling, star formation, and feedback \cite[e.g.][]{White1993,Kravtsov2005, Planelles2013, Sembolini2016a, Sembolini2016b}. In this way \cite{Eckert2019} used high-precision hydrostatic masses  out to the virial radius for a sample of 13 nearby $3 \times 10^{14}$ M$_\odot < M_{\rm 500} < 1.2 \times 10^{15}$~M$_\odot$ clusters, and found a median non-thermal pressure fraction of 6\% and 10\% at $R_{\rm 500}$ and $R_{\rm 200},$ respectively.

All of these three approaches give a fraction of non-thermal pressure consistent with subsonic turbulence with a Mach number $\approx 0.25,$ matching that of the High run as shown in Figure~\ref{fig:velocity}. Thus measurements at or above the driving scale in $\gtrsim 10^{13}$~M$_{\odot}$ systems and extrapolations down to the scale of individual absorbers in $\approx 10^{12}$~M$_{\odot}$ systems are both completely consistent with the turbulent velocities assumed in our High simulation.

\begin{figure*}
\includegraphics[width=1.00\linewidth]{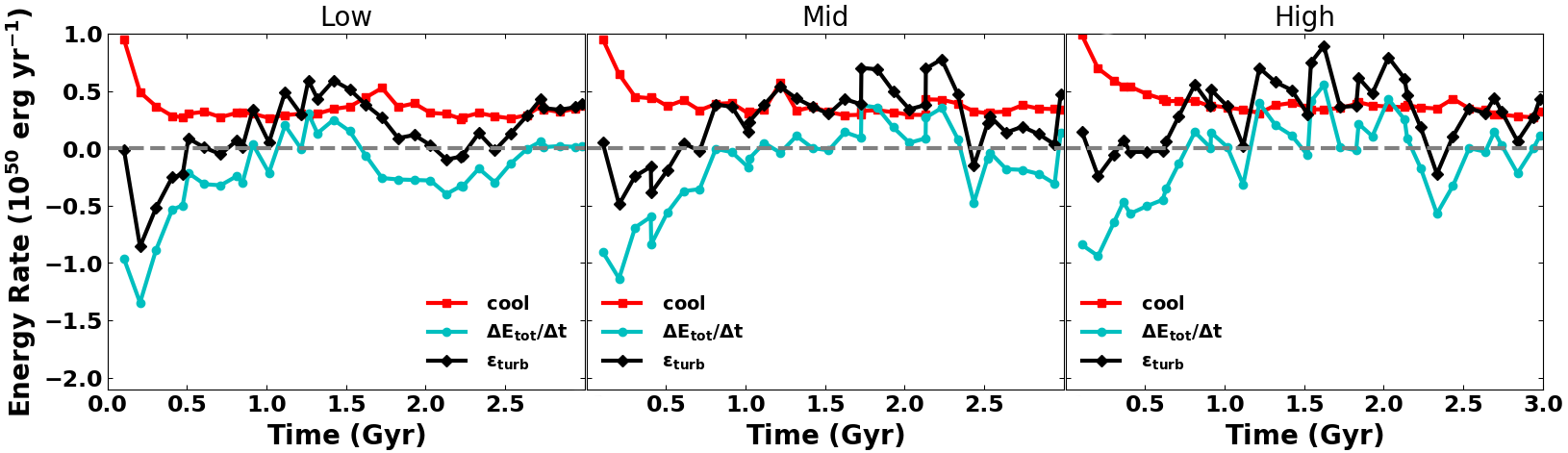}
\caption{Energy rates vs. time in cooling (red squares), $\Delta E_{\rm tot}/\Delta t$ (cyan circles), and $\epsilon_{\rm turb}$ (black diamonds) for the Low (left), Mid (middle), and High (right) runs. A grey dashed line indicates no change in energy. The energy rates are shown in units of 10$^{50}$~erg~yr$^{-1}$.}
\label{fig:energy_rates}
\end{figure*}

\subsection{Energy Requirements for CGM Turbulence} \label{energy_req}
Having shown the ionization state structure and turbulent velocities assumed in our simulations are both consistent with observations, we now investigate the amount of energy needed to sustain these motions. We find the gravitational {potential} (PE), kinetic (KE), internal (EI), and total energies (PE+KE+EI) within the virial radius for each run and show these in Figure~\ref{fig:energy}. We define these energies as the following:
\begin{gather}
    PE \equiv -4\pi G \rho_0 R_s^{3} \int dV  \frac{\rho(r)}{r} \ln   \left( 1 + \frac{r}{R_{\rm s}} \right), \\
    KE \equiv \int dV \frac{1}{2} \rho({\bf x}) v({\bf x})^{2}, \\
    EI \equiv \int dV  \rho({\bf x}) T({\bf x})  \frac{N_A k}{\left[1-\gamma({\bf x})   \right] \overline{A}({\bf x})},
\end{gather}  
with ${\overline{A}} \equiv \left(\sum_{i} X_i A_i^{-1} \right)^{-1}$, {$\rho$} and $v$ are the {density} and velocity, $X_i$ and $A_i$ are the mass fraction and atomic mass of the $ith$ species, $N_A$ is Avogadro's number, $k$ is the Boltzmann constant, $\gamma$ is the weighted average adiabatic index, {$r$ defines the distance from the center, ${\bf x}$ defines the position in space,} and the integrals are taken over the full volume within the virial radius. 

{Given the nature of our turbulence and how it is driven, it is non-trivial to back out the energy injection rate. As turbulence is driven, the energy from it is dissipated in the form of cooling as well as doing work by moving mass within the gravitational potential of the halo. Thus we find the cooling rate and rate of change in total energy, $\Delta E_{\rm tot}/ \Delta t$, the sum of which gives us the turbulent energy injection rate, $\epsilon_{\rm turb}$.} We show the cooling {rate, rate of} total energy change ($\Delta E_{\rm tot}/ \Delta t$), and {the sum of these quantities, $\epsilon_{\rm turb}$, as a function of time} in Figure~\ref{fig:energy_rates}. We also show these quantities for the double-resolution High run in Figure~\ref{fig:energy_rates_highres}. Furthermore we provide the average cooling, $\Delta E_{\rm tot}/ \Delta t$, and $\epsilon_{\rm turb}$ over 2 time periods in Table \ref{tab:energy_rates}. The first 3 rows show these rates averaged over the full simulation time while the bottom three shows the rates from 1--3~Gyrs. 

\begin{figure}
\includegraphics[width=1.00\linewidth]{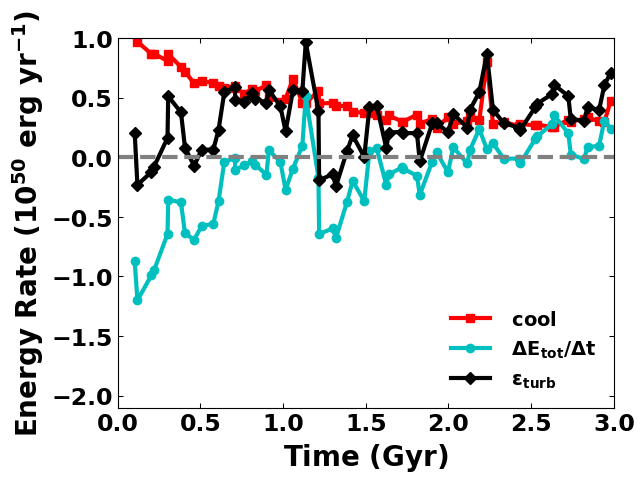}
\caption{Energy rates vs. time in cooling (red squares), $\Delta E_{\rm tot}/\Delta t$ (cyan circles), and $\epsilon_{\rm turb}$ (black diamonds) for the double-resolution High run. A grey dashed line indicates no change in energy. The energy rates are shown in units of 10$^{50}$~erg~yr$^{-1}$.}
\label{fig:energy_rates_highres}
\end{figure}

\begin{table*}[t]
\centering
\begin{tabular}{ccccc}
\hline
\hline
Run & Cooling Rate & $\Delta E_{\rm tot}/\Delta t$ & $\epsilon_{\rm turb}$ & Avg $\epsilon_{\rm turb}^{*}$ \\
& (10$^{49}$\ erg/yr) & (10$^{49}$\ erg/yr) 
& (10$^{49}$\ erg/yr) &
(SN/yr)\\
\hline
Low (t = 0--3~Gyrs) & 3.448 & -2.136 & 1.312 & 0.013 \\
Mid (t = 0--3~Gyrs) & 3.835 & -1.436 & 2.399 & 0.024\\
High (t = 0--3~Gyrs) & 4.035  & -0.883 & 3.153 & 0.032 \\
Low (t = 1--3~Gyrs) & 3.317 & -0.728 & 2.590 & 0.026 \\
Mid (t = 1--3~Gyrs)  & 3.451 & 0.263 & 3.713 & 0.037 \\
High (t = 1--3~Gyrs) & 3.475  & 0.559 & 4.034 & 0.040 \\
\hline
\end{tabular}
\caption{Average cooling, energy change ($\Delta E_{\rm tot}/ \Delta t$), and turbulent energy injection rate ($\epsilon_{\rm turb}$) for various times throughout the simulations. $^*$ shown for comparison to supernova rates with typical energy input of 10$^{51}$~ergs.}
\label{tab:energy_rates}
\end{table*}

There is a minimum $\epsilon_{\rm turb}$ below which the halo will collapse without viable pressure support and alternatively a maximum $\epsilon_{\rm turb}$ which will cause the halo to explode. Between these lies a range of energies that {will allow the halo to reach a steady-state after some time} which is why we show the rates constrained to 1--3~Gyrs and quote $4 \times 10^{49}$~erg~yr$^{-1}$, in the High run, as the amount of turbulent energy we inject per year needed to produce a similar structure and ionization state ratios as nearby star-forming galaxy halos.

For comparison, we also show $\epsilon_{\rm turb}$ in SN~yr$^{-1}$ (found by dividing by 10$^{51}$~erg) in the last column of Table \ref{tab:energy_rates} and see that the High run, in particular, requires below 5\% the kinetic energy released in a typical supernova event per year to sustain a 10$^{12}$~M$_{\odot}$ galaxy halo with many features that share similarities to the COS-Halos observations. 

We also convert our $\epsilon_{\rm turb}$ in SN~yr$^{-1}$ to a star formation rate (SFR) by finding the number of stars that are massive enough to explode in a supernova, i.e. stars with masses between 8~M$_{\odot}$ and 100~M$_{\odot}$ and multiplying by the average star mass between 0.1--100~M$_{\odot}$. Assuming a \citet{salpeter1955luminosity} initial mass function, we find that about 0.3\% of stars will experience a supernova with an average star mass of 0.4~M$_{\odot}$ for the aforementioned stellar mass range, obtaining {a SFR} of 4.8~M$_{\odot}$~yr$^{-1}$ {to supply the $\epsilon_{\rm turb}$ found in the High run}. This is comparable with the recent SFRs from the COS-Halos star-forming sample at 0.4--12~M$_{\odot}$~yr$^{-1}$ \citep{2013werkApJS..204...17W}, the Milky Way at 1.65~M$_{\odot}$~yr$^{-1}$ \citep{Licquia2015}, and 0.3--10~M$_{\odot}$~yr$^{-1}$ from the Muse Gas Flow and Wind survey of z $\approx$ 1 star-forming galaxies \citep{schroetter2019muse}. We would like to remind the reader that we inject this energy inhomogeneously following Equation \ref{equ:trend}, and that this radial trend results in the shallow entropy profile  in the High run, which allows convective motions to occur such that the gas cools while it falls towards the center.

Besides stars, turbulence can also be powered by accretion. If we consider $\epsilon_{\rm turb}$ to be fueled purely by accretion onto the dark matter halo from the IGM moving at the escape velocity of the halo, this would require an average gas mass accretion rate of 50~M$_{\odot}$~yr$^{-1}$ onto the galactic halo from the IGM. 

{For comparison we compute the actual mass inflow rate within the CGM of the High run by integrating the density times the radial velocity over a shell between 12-214~kpc. In this case, we find a mass inflow rate of $\dot{M} =$ 29~M$_{\odot}$~yr$^{-1}$ at 3~Gyrs (24 M$_{\odot}$~yr$^{-1}$ for the double-resolution High run) which is similar to the solution for gas between 10-200~kpc  \citet{stern2019cooling}. This is also close to other theoretical work which give a gas accretion rate of $\approx 10-20$~M$_{\odot}$~yr$^{-1}$ for $10^{12}$~M$_\odot$ halos at $z=1$ to 2 \citep[e.g.][]{vandeVoort2011}. If we change the limits of integration to12-14~kpc, to study the inter halo, we obtain a mass inflow rate of 4~M$_{\odot}$~yr$^{-1}$. Interestingly, the mass inflow rate found in the \citet{stern2019cooling} model that produced a good match to the higher O ions was $\dot{M} =$ 1.65~M$_{\odot}$~yr$^{-1}$, which is close to our results and also matches the SFR of the Milky Way.}

\begin{figure}
\includegraphics[width=1.00\linewidth]{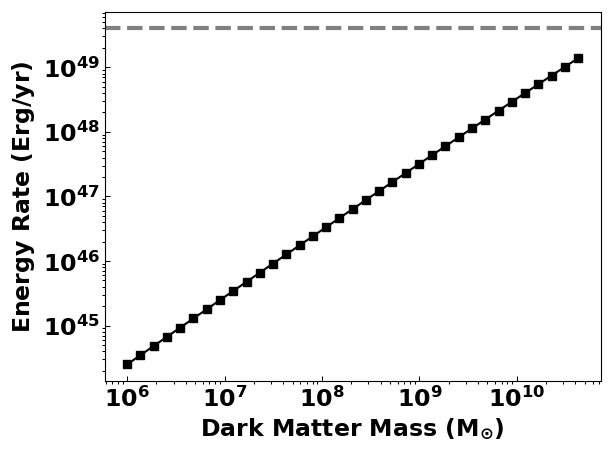}
\caption{Total Energy loss rate vs. dark matter subhalo mass with $\epsilon_{\rm turb}$ shown as a grey dashed line.}
\label{fig:subhalo}
\end{figure}

Finally, we consider the energy loss rate that would result from the gravitational drag of dark matter subhalos moving through the ambient dark matter density of the galactic halo. To estimate this we convert the ambient baryonic density of 10$^{-27}$~g~cm$^{-3}$ to an ambient dark matter density and assume subhalos are traveling at $V$ = 141~km~s$^{-1}$, the dispersion velocity of the galactic halo at the virial radius. This being the subsonic case (the average sound speed at 3~Gyrs is $c_s$ = 160~km~s$^{-1}$), we approximate the energy loss rate as: 
\begin{equation}
    -\frac{dE}{dt}=\frac{(96\pi)^{1/2} G^2 M_{sh}^2 \rho_0 V^2}{c_s^2} \ln \Lambda,
\end{equation}
where  $\rho_0$ is the ambient gas density of the galactic halo, $M_{sh}$ is the mass of the subhalo, $V$ is its velocity, $\Lambda$ is $R_{\rm vir}/r$, and $r$ is the radius of the dark matter subhalo \citep{Chandrasekhar1943,rephaeli1980flow}.
Combining this with the dark matter mass function from \citet{springel2008aquarius} gives the  integrated energy loss rate from all subhalos below a given mass:
\begin{equation}
    -N\frac{dE}{dt}=(1.85 \times 10^{45}~g)^{0.9} \frac{\rho_0 G^2 M_{sh}^{1.1} v^2}{c_s^3} \ln \Lambda.
\end{equation}
We consider subhalos with masses between 10$^6$--$4.4 \times 10^{10}$~M$_{\odot}$, the upper limit being the mass beyond which we have only fractions of a subhalo. We find $r$ by assuming the subhalos have the same density as the central galactic dark matter density, $3.67 \times 10^{-25}$~g~cm$^{-3}$, as well as a spherical shape yielding radii ranging from 350 -- 12,000~pc. Finally, we show the energy loss rate from all subhalos at a given mass as a function of subhalo mass in Figure \ref{fig:subhalo} and see that this can reach levels as high as about a third of the overall energy input rate needed.

In reality it is likely a mixture of supernovae, gas inflows, and dynamical friction from dark matter subhalos that work together to drive turbulence in galaxy scale halos.  These in turn help to drive convective flows, which  produce large scale  cold filaments and clouds that are surrounded by layers of rapidly cooling gas, contained within a hot diffuse medium.

\section{Discussion and Summary} \label{summary}

Motivated by recent COS-Halos results, we have used our updated MAIHEM model to carry out a suite of chemodynamical simulations of turbulent media in a Navarro-Frenk-White (NFW) gravitational potential that match the circumgalactic media of low redshift galaxies.  In all runs, turbulence was driven on scales between 10 to 30~kpc, with stirring that was strongest toward the center and fell off gradually with radius. We looked at 3 cases for the average $\sigma_{1D}$: 20, 34, and 41~km~s$^{-1},$ which we labeled as the Low, Mid, and High run, respectively.

 All three turbulent runs maintain the halo equilibrium temperature throughout their 3~Gyrs runtimes. While the Low run shows a strong entropy gradient and limited inhomogeneities, the High run shows a flatter entropy profile that sets up convective motions throughout the halo. These motions in turn produce an inhomogeneous medium that contains low and intermediate ionization state material at large radii. 

Comparing the projected column densities of a wide range of ions with those from the COS-Halos survey, we found that all three runs can reproduce the observations at low impact parameters, but only the High run has sightlines at large impact parameters that overlap with the COS-Halos detections and limits. We also find good agreement between our \ion{O}{6} columns and those from other recent theoretical and observational work. Furthermore, unlike equilibrium models, our simulations produce many sightlines with $n \leq$ 10$^{-4}$~cm$^{-3}$ that match the $N_{\rm Si\ IV}$/$N_{\rm O\ VI}$ and $N_{\rm N\ V}$/$N_{\rm O\ VI}$ limits observed in the COS-Halos data. The suppressed ratio arises from gas that cools even faster than isochorically, due the the combination of radiative cooling and the turbulent mixing of hot and cold media.

When comparing to more recent results from \citet{lochhaas2019high}, we find our $N_{\rm N\ V}$/$N_{\rm O\ VI}$ in agreement with some of the detections and limits, however, we find no sightlines that agree with their lowest upper limits on this ratio.

The turbulent energy injection rate in our High simulation, $\epsilon_{\rm turb} \approx 4 \times 10^{49}$~erg~yr$^{-1}$, is comparable to expectations from several energy generation mechanisms that may be present in the CGM of low redshift star-forming galaxies such as supernovae, mass accretion, and stirring by dark matter subhalos. In particular $\epsilon_{\rm turb}$ is consistent with supernovae arising from a star formation rate of $\approx 5$~M$_{\odot}$~yr$^{-1}$, or a mass infall rate of 50~M$_{\odot}$~yr$^{-1}$ onto the galactic halo. Dark matter subhalos, however, can contribute significantly but are unlikely to be able to supply this energy rate by themselves. 

We acknowledge that low ions are typically observed with lower velocity dispersions than the higher ions \citep{tuml2013ApJ...777...59T,churchill2015direct,werk2016ApJ...833...54W} and by design our simulations had a higher velocity dispersion near the center of the halo, where low ionization state ions reside. However, we also found that on the smallest scale of our simulations, the internal turbulence within individual low ionization absorbers is comparable to those found in the observations.

Lastly, we conducted a second High run with double the resolution of our other runs and found it to have features similar to a mix between the original Mid and High runs. In this case, the gas cools more efficiently towards the center for the beginning of the simulation, setting up a CGM that is slightly less dense and overall hotter than that of the original High run, but still subject to large-scale convection. This facilitates the production of smaller structures in place of filaments, but the results remain in agreement with observations of the outer CGM, and the ion ratios also overlap the observed $N_{\rm Si\ IV}$/$N_{\rm O\ VI}$ and $N_{\rm N\ V}$/$N_{\rm O\ VI}$ ratios. {More resolved cooling alters the non-equilibrium chemistry at 3~Gyrs only slightly as compared to the original resolution High run, with all of the elements varying by $\lesssim$ 3\% except for  Fe, which varies by $\approx 8$\% as compared to the original resolution simulation.}

Future work will include adding more ions to our chemistry network, varying the radial profile of the turbulence, and studying the impact of rotation and electron thermal conduction. This will help yield a more complete understanding of the kinematic structure of the multiphase medium around star-forming galaxies. 

\acknowledgments
We would like to thank Jessica Werk for providing the COS-Halos data shown, Sanchayeeta Borthakur for her advice on absorption lines measures of low-redshift galaxy halos, and Hsiao-Wen Chen, Drummond Fielding, and Michele Fumagalli for providing useful comments and feedback. Helpful comments by the referee are also gratefully acknowledged. E.B.II was supported by the National Science Foundation Graduate Research Fellowship Program under grant No. 026257-001. E.S. gratefully acknowledges the Simons Foundation for funding the workshop Galactic Winds: Beyond Phenomenology, which helped to inspire this work and the the Max Planck Society for funding the CGM Berlin 2019 workshop, which led to discussions that greatly improved it.  E.S. was supported by NSF grant AST14-07835 and NASA theory grant NNX15AK82G. The simulations presented in this work were carried out on the Stampede2 supercomputer at Texas Advanced Computing Center (TACC) through Extreme Science and Engineering Discovery Environment (XSEDE) resources under grant TGAST130021.\\

\textit{Software}: FLASH \citep[v4.5]{fryxell2000flash}, Cloudy \citep{ferland20132013}, yt \citep{2011turk}

\bibliographystyle{yahapj}
\bibliography{references}

\appendix


\section{Column Density Maps \& Profiles}
\label{app:colden}
In this section we show our projected column density maps for \ion{C}{2}, \ion{C}{3}, \ion{Mg}{2}, \ion{N}{2}, \ion{O}{1}, and \ion{Si}{4}. We also show these projected column density maps for \ion{Si}{2}, \ion{Si}{3}, \ion{N}{5}, \ion{O}{6}, \ion{C}{2}, \ion{C}{3}, \ion{Mg}{2}, \ion{N}{2}, \ion{O}{1}, and \ion{Si}{4} from the double-resolution High run. In addition to this, we show the column density profiles for the aforementioned ions as well as the temperature and \ion{H}{1} column density profiles with data from the double-resolution High run. 

\begin{figure*}[h]
\centering
\includegraphics[width=1.00\linewidth]{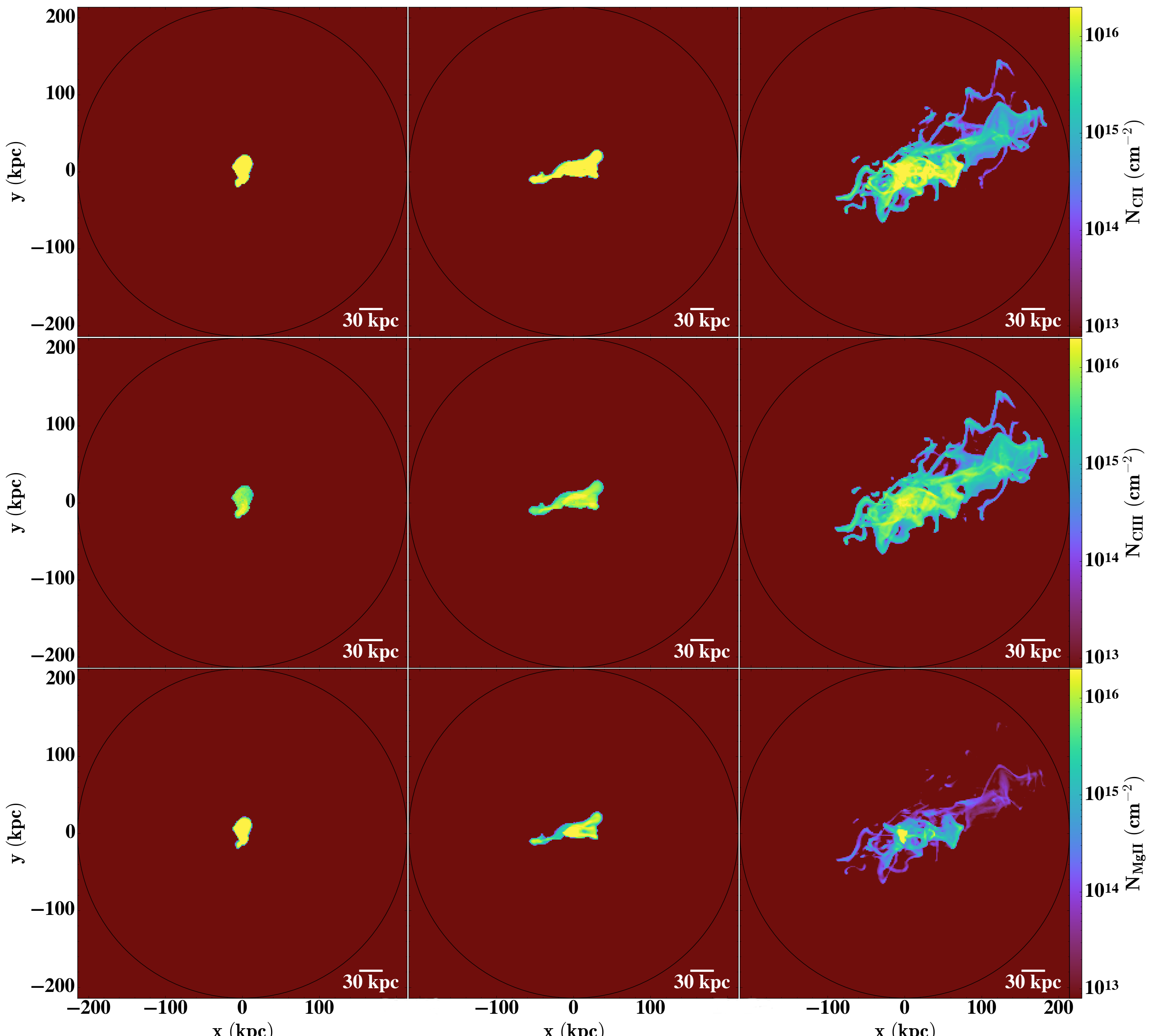}
\caption{$N_{\rm C\ II}$ (first row), $N_{\rm C\ III}$ (second row), and $N_{\rm Mg\ II}$ (third row) projections for the Low (left column), Mid (middle column), and High (right column) runs at 3~Gyrs. A black ring shows the virial radius at $r \approx 220$~kpc.}
\label{fig:project1}
\end{figure*}

\begin{figure*}
\centering
\includegraphics[width=1.00\linewidth]{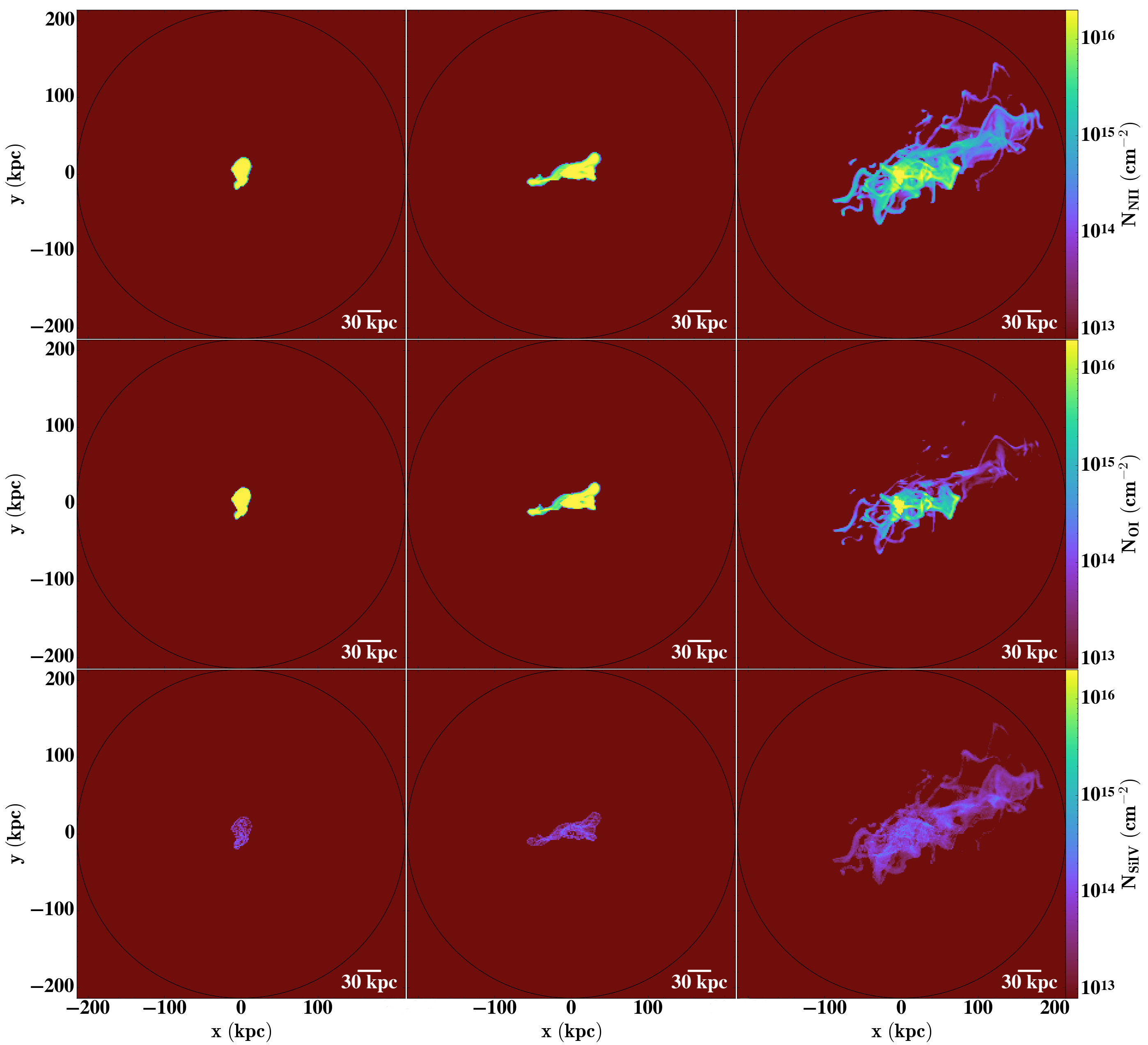}
\caption{$N_{\rm N\ II}$ (first row), $N_{\rm O\ I}$ (second row), and $N_{\rm Si\ IV}$ (third row) projections for the Low (left column), Mid (middle column), and High (right column) runs at 3~Gyrs. A black ring shows the virial radius at $r \approx 220$~kpc.}
\label{fig:project2}
\end{figure*}

\begin{figure*}[h]
\centering
\includegraphics[width=1.00\linewidth]{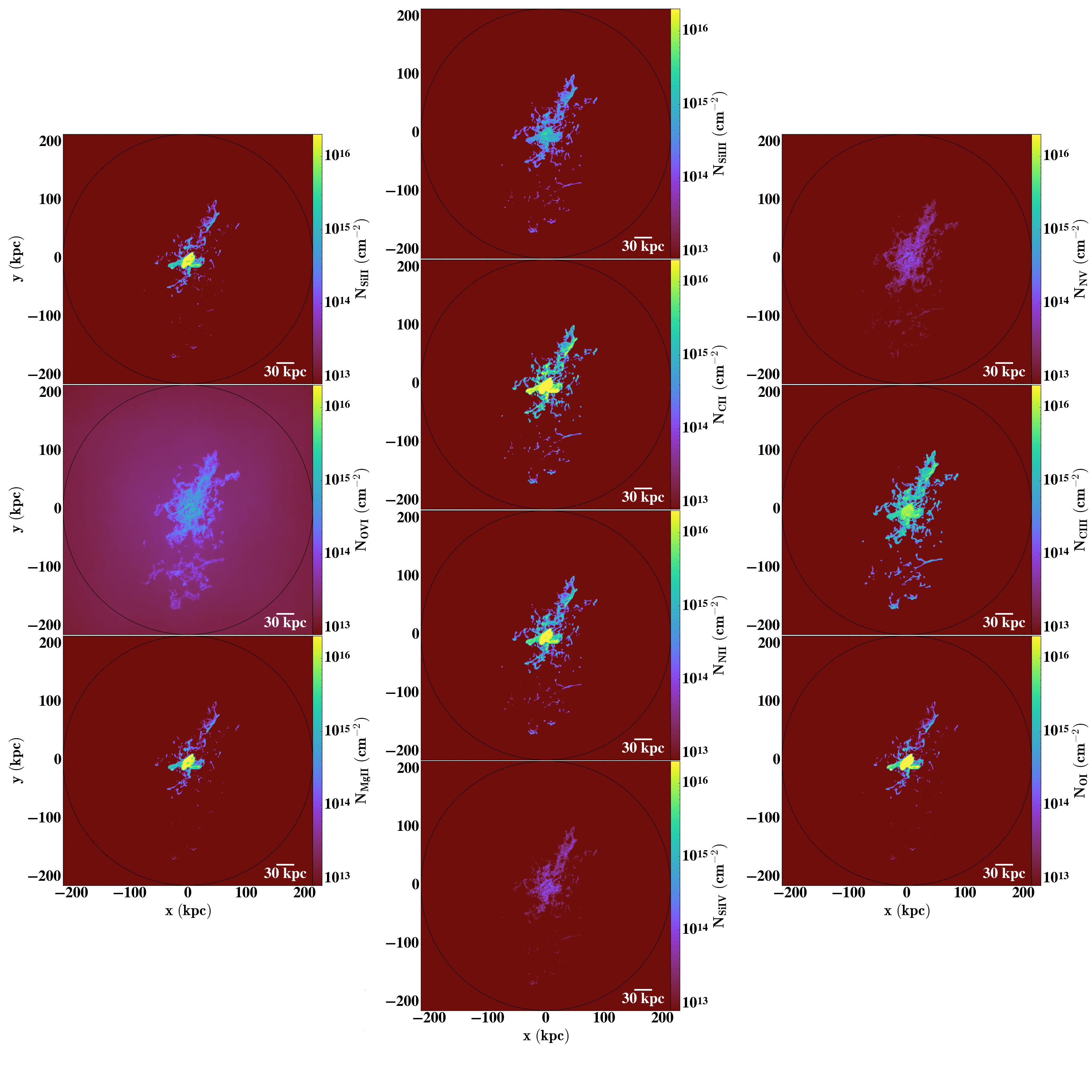}
\caption{$N_{\rm Si\ II}$ (top, first column), $N_{\rm Si\ III}$ (top, second column), $N_{\rm N\ V}$ (top, third column), $N_{\rm O\ VI}$ (middle, first column), $N_{\rm C\ II}$ (second, second column), $N_{\rm C\ III}$ (middle, third column), $N_{\rm Mg\ II}$ (bottom, first column), $N_{\rm N\ II}$ (third, second column), $N_{\rm O\ I}$ (bottom, third column), and $N_{\rm Si\ IV}$ (bottom, second column) projections for the double-resolution High run at 3~Gyrs. A black ring shows the virial radius at $r \approx 220$~kpc.}
\label{fig:highres_maps}
\end{figure*}

\begin{figure*}
    \centering
    \includegraphics[width=1.0\linewidth]{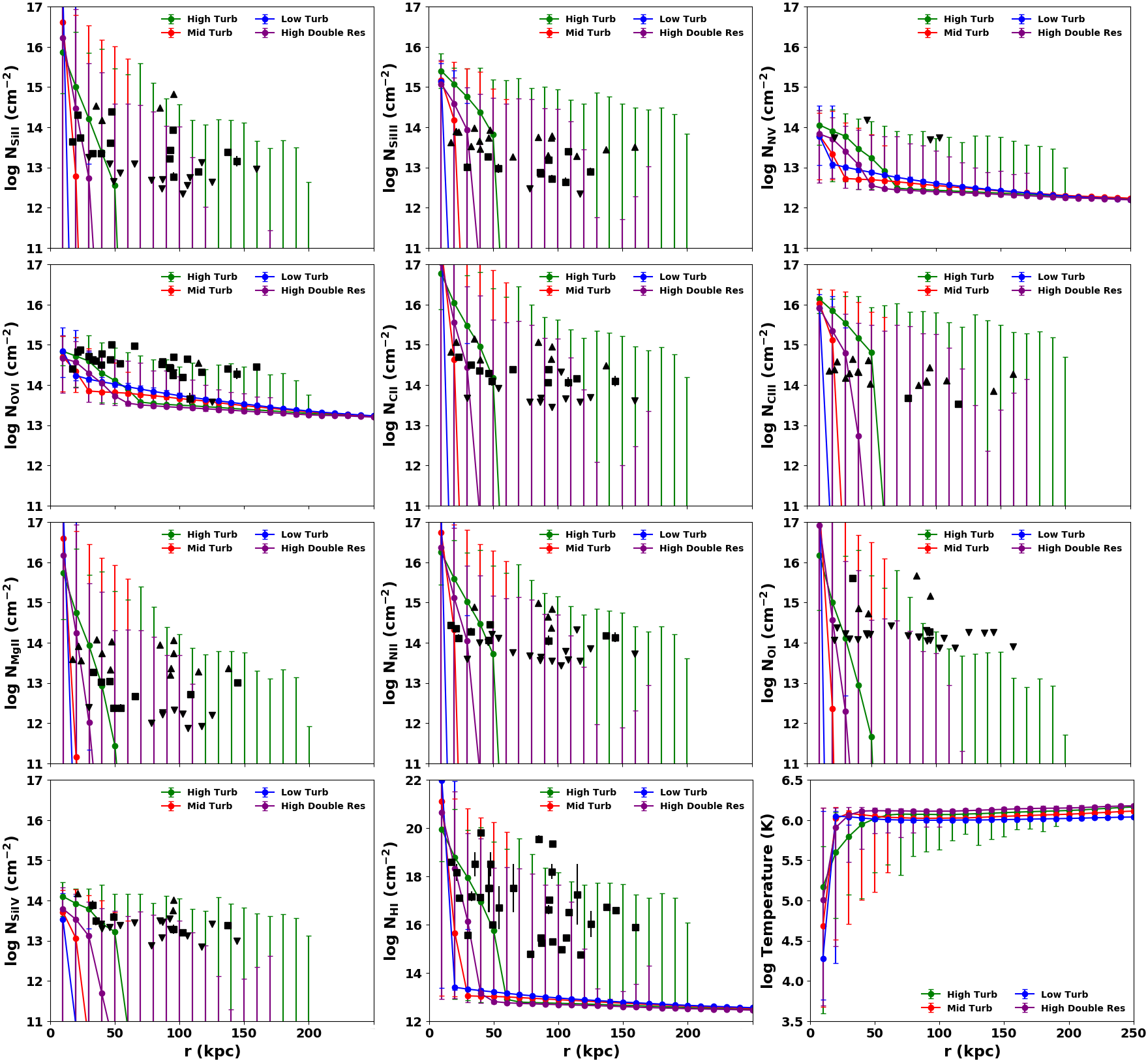}
    \caption{{Filled circles show the 50$^{\rm th}$ percentile of sightlines in 10~kpc bins} for Log $N_{\rm Si\ II}$ (first row left), $N_{\rm Si\ III}$ (first row middle), $N_{\rm N\ V}$ (first row right), $N_{\rm O\ VI}$ (second row left), $N_{\rm C\ II}$ (second row middle), $N_{\rm C\ III}$ (second row right), $N_{\rm Mg\ II}$ (third row left), $N_{\rm N\ II}$ (third row middle), $N_{\rm O\ I}$ (third row right), $N_{\rm Si\ IV}$ (fourth row left), $N_{\rm H\ I}$ (fourth row middle), and Temperature (fourth row right) vs. radial distance $r$ for the Low (blue), Mid (red), High (green), and double-resolution High (purple) runs at 3~Gyrs contained to cells within the virial radius. {Error bars show the 1$^{\rm st}$ percentile as the lower limit and 99$^{\rm th}$ percentile as the upper limit.} Projections were generated for each cell where each sightline is a line integral of ion number density along the z-axis. We also show detections (black squares) and limits (black arrows) from the COS-Halos sample overlaid on our projected column densities.}
    \label{fig:highres_columns}
\end{figure*}

\end{document}